\documentclass[12pt,onecolumn]{IEEEtran} %
\usepackage{graphics} 
\usepackage{graphicx}
\usepackage{verbatim}
\usepackage{xcolor}
\usepackage{mathtools}
\usepackage{amsmath}
\usepackage{amsthm}
\usepackage{amsfonts}
\usepackage{mathtools}
\usepackage{cancel}

\usepackage[ruled,linesnumbered]{algorithm2e}
\usepackage{float}
\usepackage{subfig}

\usepackage{hyperref}
\hypersetup{
	colorlinks=false,
	urlbordercolor={0 0 1},
	pdfborderstyle={/S/U/W 1},
}

\newcommand{\remark}[1]{ {\color{red}  \sc ($\longleftarrow$ #1)} }

\newtheorem{theorem}{Theorem}
\newtheorem{lemma}{Lemma}
\newtheorem{corollary}{Corollary}
\newtheorem{definition}{Definition}

\def\conv{\otimes}

\begin{document}

\title{Network Calculus Characterization of Congestion Control for Time-Varying Traffic}

\author{Harvinder Lehal,~Natchanon Luangsomboon,~J\"{o}rg Liebeherr
\thanks{
The authors are with the Department of Electrical and Computer Engineering, University of Toronto. 
}}

\maketitle
\begin{abstract}
Models for the dynamics of congestion control  generally involve systems of coupled  differential equations. Universally, these models 
assume that traffic sources saturate the  maximum transmissions 
allowed by the congestion control method. This is not suitable for studying  congestion control  
of intermittent  but bursty traffic sources. 
In this paper, we present a characterization of congestion control for arbitrary time-varying traffic that applies to rate-based as well as window-based congestion control. We leverage the capability of network calculus to precisely describe the input-output relationship at network elements for arbitrary source traffic.  We show 
 that our characterization can closely track the dynamics of even complex congestion control algorithms. 
\end{abstract}



\section{Introduction}
\label{sec:intro}

Congestion control in data center networks  must deal with highly dynamic workloads with low latency requirements that  exchange massive amounts of traffic. 
A particular difficulty arises from short-lived traffic surges, known as microbursts, that cause periods of  high packet delay and loss~\cite{microburst1,microburst2,microburst6,microburst8}. Since microbursts occur on a scale below that of a round-trip time, traditional congestion control algorithms (CCAs) are often not well equipped for such scenarios. 
As a consequence, the quest for effective congestion control in data centers has diverged from congestion control research for long-haul networks~\cite{DCQCN-Sigcomm2015,dctcp,HPCC-Sigcomm19,timely,Swift-Sigcomm20,Mushtaq2019,Ganjali22,microburst4,microburst7}. 

The performance and effectiveness of CCAs for data centers can (and generally is)  evaluated in measurement experiments on testbed networks or packet-level simulations. Since measurement studies are tied to the concrete 
network and protocol settings in which experiments are conducted, 
generalizations and extrapolations of outcomes can be difficult.  A comparable evaluation of CCAs through measurement experiments  is not an option when different methods are implemented in different protocol stacks. This is an acute issue in data centers where TCP has been increasingly replaced by RDMA over Converged Ethernet (RoCE) as transport  protocol~\cite{infiniband,rocev2}. While some, especially earlier, congestion methods for data centers are integrated into TCP, e.g., DCTCP~\cite{dctcp}, AC/DC TCP~\cite{acdc-tcp}, HCC~\cite{Ganjali22}, 
more recent proposals, such as 
DCQCN~\cite{DCQCN-Sigcomm2015}, Timely~\cite{timely}, HPCC~\cite{HPCC-Sigcomm19}, and Swift~\cite{Swift-Sigcomm20},  involve RDMA transport. 
A performance evaluation of CCAs using a model-based approach can enhance the insights obtained from measurement experiments. A significant benefit of a model-based analysis is the ability to characterize CCAs at a more abstract level, thereby providing a clearer understanding of their  operational dynamics and properties. 
A  concern with a model-based analysis of CCAs is the accuracy at which underlying protocols are  represented. 

The classic model-based analysis approach for CCAs employs a fluid model that expresses  network dynamics in terms of a coupled system of delay differential equations~\cite{fluid-towsley-03,fluid-towsley-21,Book-srikant,fluid-low-02}. 
Here, different aspects of a CCA, such as the congestion window, round-trip time, and packet loss rate, are represented as variables, and differential equations describe 
how the change of one variable affects others. For example,  increasing the  congestion window leads to higher throughput, but it also increases the likelihood of packet loss.
Fluid models for studying congestion control are well-explored~\cite{fluid-srikant-02} and remain the preferred method for modelling the dynamics of CCAs~\cite{DCQCN-Sigcomm2015,DCTCPanalysis-11,fluid-BBR-22}.

While models based on  differential equations provide powerful tools   
for evaluating convergence and stability properties of a CCA, they universally assume that traffic sources are saturated in the sense that they consistently transmit at the maximum 
rate set by the CCA. Such models are not suitable for 
evaluating  CCAs in traffic scenarios that are susceptible to microbursts, 
that is, short-lived traffic surges. 
A notable example of a traffic class exhibiting these characteristics is distributed training of 
deep neural networks. Here, the traffic of workers involved in  the training follows an on-off pattern, where  time periods with 
few transmissions alternate with time periods featuring a sequence of burst transmissions.\footnote{The number of bursts relates to 
the number of convolution layers of the neural network and the burst sizes are determined by the number of parameters within  those layers.} Even a relatively modest model such as ResNet50~\cite{resnet50} generates  up to 10~MB per bursts, with an average 
transmission rate of 300--400~Mbps per worker in a computation-limited scenario.\footnote{In a computation-limited distributed training scenario, the gradient computation is slower than the communication during gradient reduction.}

For studying CCAs under intermittent but bursty traffic, we take a different approach, where we describe congestion control algorithms within the framework of the network calculus~\cite{Book-LeBoudec,Book-Chang,Book-Bouillard}. Taking advantage of the min-plus convolution operation 
of the network calculus, we are able to study CCAs for arbitrary time-varying traffic patterns.  Our  network representation is inspired by the path server model from~\cite{Arun_2021}, 
which takes a network calculus inspired approach for model checking of CCA loss scenarios. 
We show that a network calculus characterization can closely track the dynamics of even complex congestion control algorithms. 
In this paper, we make the following key contributions:
\begin{itemize}
    \item In \S\ref{subsec:events}, we express events relevant to a CCA, such as acknowledgements, timeouts, retransmissions,  congestion notifications, and round-trip times, within the formalism of the network calculus. 
    \item In \S\ref{subsec:rate}~and~\S\ref{subsec:window}, respectively, we present models that characterize the dynamics of rate-based and  window-based CCAs. By considering time intervals where the network satisfies min-plus linearity and shifting  the coordinate system between intervals, we are able to perform a linear analysis of an overall non-linear network system. A comparison with simulations for TCP Vegas~\cite{Vegas} evaluates the accuracy of the model.
    \item In \S\ref{sec:multiflow}, we address congestion control for flows that share a bottleneck resource. We derive  a new network calculus result, which reconstitutes individual traffic flows at the egress of a multiplexer. With this result, we show that our network calculus expression captures convergence toward fairness. 
    \item \S\ref{sec:casestudy} presents a case study of a burst transmission scenario, where we address the interaction between commonly deployed  flow control and congestion control methods in data centers with RDMA transport. The case study shows that a network calculus characterization is able to closely track a packet-level simulation.
\end{itemize}

\section{Network Calculus Background}
\label{subsec:netcalc}
Network calculus~\cite{Book-LeBoudec,Book-Chang,Book-Bouillard} is a methodology for characterizing the input-output relationship of 
traffic at a network element, where a network element may represent one or a 
collection of switches, traffic control algorithms, or transmission media. Different from pure fluid flow models,   traffic may arrive  to and depart from network elements in bursts. As a result, the description of cumulative traffic is not continuous. 
We describe  traffic arriving to and departing from a network element by left-continuous functions $A$ and~$D$, such that 
$A(t)$ and $D(t)$, respectively, represent the cumulative arrivals and departures at a network element in the interval $[0,t)$,  with $D(t)\le A(t)$. 
The service at a network element  is described by another function $S$,  referred to as the element's service curve. 
Like functions $A$ and $D$, a service curve $S$ is non-negative and non-decreasing, and we set $A(t) = D(t) = S(t) =0$ for $t \le 0$. 
The backlog at a network element $B$ is a function that describes the traffic arrivals that 
have not departed, that is, $B(t) = A(t)-D(t)$. 

A network element transforms  arriving traffic to generate the traffic pattern that departs the network element. With network calculus, the interaction between arriving traffic and a network element is described in terms of a min-plus convolution of the arrival function and the service curve of the network element. The min-plus convolution  of two functions $f$ and $g$, denoted by $f \conv g$, is defined by $f\conv g (t) = \inf_{0 \le s \le t} \{f (s) + g (t-s)\}$.\footnote{With non-continuous functions, we must use infimums and supremums, respectively, instead of minimums and maximums.}  If $D (t) = A \conv S(t)$ holds for all times~$t$, the service curve $S$ is referred to as an exact service curve. If the inequality $D (t) \ge A \conv S(t)$ or $D (t) \le A \conv S(t)$ holds for all~$t$, we speak of a lower or upper service curve, respectively \cite{Book-LeBoudec}. 
An egress port with link rate $r >0$ can be characterized by an exact service curve 
$S(t) = \max\{rt, 0\}$, and a token bucket with bucket size $b>0$ and 
rate $r>0$ has an exact service curve $S(t) = \max \{b + r\, t , 0\}$.

Networked systems that can be described by exact service curves are min-plus linear, in the sense that they are time-invariant and satisfy the superposition principle using the operations of a  min-plus dioid algebra \cite{Book-LeBoudec}. 
While there exist feedback systems that satisfy min-plus linearity~\cite{Cruz99},  delayed feedback  and other factors of congestion control result in a system that is not min-plus linear. 
In the network calculus,  systems that are not min-plus linear can be   characterized by lower and upper service curves, however, at a loss of accuracy. In this paper, we 
maintain a min-plus linear characterization by  dividing the time axis into intervals in which the network system satisfies min-plus linearity  and characterizing the traffic separately for each interval. 

As a remark, the fact that network calculus lends itself to a worst-case analysis 
of networks (involving lower service curves and upper bounds for arrivals) 
has invited a  misconception that network calculus computations are generally pessimistic. 
We note that a characterization of min-plus linear systems with exact service curves yields exact quantities for departures, backlog, etc..

We sometimes need to compute the horizontal distance between~$A$ and $D$, or similar functions. If these functions were continuous and strictly increasing, we could compute the horizontal distance between $A$ and $D$ at time $t$ as $t - A^{-1} (D(t))$, where $A^{-1}$ is the inverse function of~$A$ with~$A^{-1} (A(t)) = t$. However, due to burst arrivals and departures, the  functions $A$ and $D$  are not  continuous and may have plateaus where no arrivals or departures occur for some time. For these functions, we can compute the horizontal distance 
using pseudo-inverse functions~\cite{Liebeherr17,Zippo23}. For a non-decreasing function $F$, the upper pseudo-inverse $F^\uparrow$ and lower pseudo-inverse $F^\downarrow$ are defined as 
\[
F^\uparrow (y) = \sup \{ x \, \mid \, F(x) \le y \}   \ \ \text{ and }  \ \ 
F^\downarrow (y) = \inf \{ x \, \mid \, F(x) \le y \}  \, .  
\]
With this, the horizontal distance between $A$ and $D$ at time $t$ can be computed with an upper pseudo-inverse as $t - A^{\downarrow} (D(t))$ or $D^{\downarrow} (A(t)) - t$, depending on the view point.

\section{Network Calculus Model for Congestion Control}

Our characterization of congestion control techniques exploits that a network that is not min-plus linear may satisfy min-plus linearity within finite  intervals. This suggests an iterative approach where we describe 
the network behavior by an exact service curve for as long as 
min-plus linearity holds. Then, we 
establish a new reference system that exploits min-plus linearity in the 
next time interval. 

\begin{figure}[t!]
    \centering
    \includegraphics[width=\textwidth]{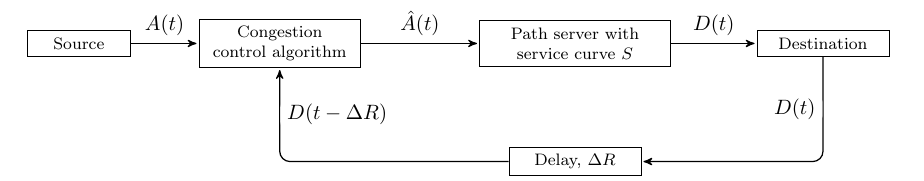}
     \caption{Path server model with congestion control.}\label{fig:NetworkCalculusModel}
\end{figure}

\subsection{Single Flow Model}

For a single flow, we resort to the network model shown in  Figure~\ref{fig:NetworkCalculusModel}, 
which is inspired by the path server model from~\cite{Arun_2021}. 
We use the following notation for traffic functions:
\begin{center}
\begin{tabular}{r l l}
$A (t)$ --& Arrival function & -- \ Source traffic generated in the interval $[0,t)$, \\
$\hat{A}(t)$ --& Admitted arrival function &-- \ Traffic admitted by a CCA in $[0,t)$, \\
$D (t)$ --& Departure function &--  \ Traffic that is received by the destination in $[0,t)$,\\
\end{tabular}
\end{center}
with $A (t) \ge \hat{A}(t) \ge D(t)$ for all $t$. 
In Figure~\ref{fig:NetworkCalculusModel}, a source generates traffic according to arrival function~$A(t)$. A congestion control algorithm filters the source traffic and admits $\hat{A}(t) \le A(t)$ into the network. This traffic  enters a path server, with exact service 
curve~$S(t)$. The traffic arriving at the destination,~$D(t)$, 
can then be obtained using the min-plus convolution
\[
D(t) = \hat{A} \otimes S (t).
\]
In the path server model from~\cite{Arun_2021}, the network service is 
described by a token bucket variant, where tokens represent the network capacity available to a traffic flow of interest. In our model, the path server is an arbitrary exact service curve.

\subsection{Modelling Congestion Events}
\label{subsec:events}
CCAs react to or trigger various congestion events that update the state of the algorithm. We model these signals by analyzing the interaction between arrival and departure functions. 

\paragraph{Acknowledgments}
For each arrival at the destination, we assume that the destination sends acknowledgements back to the source. Assuming, as in the path server model of ~\cite{Arun_2021}, a fixed delay~$\Delta R$, the cumulative acknowledgement function $D(t - \Delta R)$  expresses the arrival of 
acknowledgments at the source. Assuming that an acknowledgement is issued whenever the equivalent of the maximum packet size~$I_{\max}$ has been transmitted, a packet-level version of the acknowledgment function is given by the step function 
\[ 
I_{\max} \left\lfloor \frac{D(t-\Delta R)}{I_{\max}}\right\rfloor \, . 
\]

\paragraph{Timeouts}
Packet timeouts are used by some CCAs to indicate that the network is facing congestion and that the transmission rate should be lowered. In TCP, this is referred to as an RTO timeout~\cite{Floyd_1996}. Let $\tau_o$ represent the maximum duration that a packet can be unacknowledged. To simplify the analysis, we assume that $\tau_o$ is a constant. We can represent the least amount of traffic that should be acknowledged by time $t$ as $\hat{A}(t-\tau_o)$, and refer to it as the \emph{timeout function}. A timeout occurs at 
time~$t$, when the acknowledgements function falls below the timeout function, that is,  
\[
D(t-\Delta R)<\hat{A}(t - \tau_o) \, .
\]

\paragraph{Retransmissions}
Timeout congestion events may result in retransmissions of data, generally by resuming transmissions starting at the data that triggered a timeout (Go-Back-N). 
Since traffic functions for arrivals and 
departures in the network calculus are assumed to be non-decreasing, retransmissions do not fit easily into the framework. In our analysis, this is addressed through a shift of  the coordinate system, whereby the origin is moved to the coordinates of the first retransmission. The details of this technique are given in \S\ref{subsec:rate}.

\paragraph{Explicit Congestion Notification (ECN)}
ECN \cite{ECNrfc} is a method for signaling network congestion before a packet loss has occurred. When the backlog at a switch buffer exceeds a threshold $K_{\max}$, packets are marked with a congestion signal. When such a signal arrives at the destination, a congestion notification is sent back to the source, which then reduces its transmissions.  ECN methods often incorporate  Random Early Drop (RED)~\cite{RED}, which adds a   second buffer threshold $K_{\min}$~($K_{\min} < K_{\max}$) and a probability $P_{\max}$ ($0< P_{\max} \le 1 $), and  marks packets that 
encounter a backlog of $x$ with probability  $p(x)$ given 
by 
\[
p(x) = 
\begin{cases}
1 \, , & {\rm if } \, x\geq K_{\max} \, , 
\\
P_{\max}\frac{x-K_{\min}}{K_{\max}-K_{\min}}\, , &{\rm if } \, K_{min} < x < K_{max} \, , \\
0\, , &{\rm if } \, x \le  K_{max}  \, . \\
\end{cases}
\]
CCAs generally  issue or react to congestion events at most once per round-trip time. This can be enforced either through  timers, as in DCQCN~\cite{DCQCN-Sigcomm2015} or by counting bytes as in TCP~\cite{ECNrfc}.

We describe an ECN variant that specifies a minimum time interval $\Delta \tau_{\rm ecn}$  between congestion notifications. 
In the single flow model, the backlog, $B$, in the path server at time $t$ is given by 
\[
B(t) = \hat{A}(t) - D(t) \, , 
\]
and a packet is marked at time $t$ if $B(t) \ge K_{\rm max}$. 
We assume that congestion notifications can be triggered by a destination 
as soon as the packet is marked. Every notification arrives at the source with a fixed latency $\Delta  R$. 

Suppose a packet is marked at time $t$ ($B(t) \ge K_{\rm max}$). Let $\underline{t}$ denote the transmission time of the most recent notification 
before $t$ or, in case a notification has been scheduled for a time after time $t$, the next notification after time $t$. There are three cases to consider: 
\begin{enumerate}
    \item $t-\underline{t} \ge \Delta \tau_{\rm ecn}$: In this case, the notification can be sent immediately at time $t$, and we set $\underline{t}=t$. 
    \item $0 \le t-\underline{t} \le  \Delta \tau_{\rm ecn}$: In this case, the next notification is sent at  $\underline{t} + \Delta \tau_{\rm ecn}$, and we set  $\underline{t}=\underline{t} + \Delta \tau_{\rm ecn}$. 
    \item $t-\underline{t} < 0$: Here a notification has already been scheduled in the future, and no action needs to be taken. 
\end{enumerate}
To make sure that the first time $t$ with $B(t)  \ge K_{\rm max}$ triggers a congestion notification, we initially set $\underline{t}= -\infty$. 
To account for RED, the marking of  traffic must additionally consider backlog in the range $[K_{\min}, K_{\max})$ where traffic is marked  according to the above probability function. 

\paragraph{Priority Flow Control (PFC)}
Specified in~\cite{pfc}, PFC  
is a link-level flow control mechanism  between two switches or between a switch and a host. 
PFC seeks to achieve a lossless service by stopping arrivals from upstream switches or hosts. 
The endpoints of a link are designated as sender or receiver.  
A backlog exceeding threshold $X_{\rm off}$ at an ingress port of a receiver  indicates that a link is congested. 
In this case, the receiver temporarily stops 
additional arrivals from the sender by sending it a  \emph{Pause} packet. By pausing transmissions sufficiently early, buffer overflows at the receiver can be fully prevented. 
Transmissions by the sender are paused until the backlog at the receiver reaches~$X_{\rm on}$~($X_{\rm on} < X_{\rm off})$.  
Assuming again a constant delay $\Delta R$ for the {\it Pause} packet in the path server model, the next time after time $t$ when transmissions are paused,  
is given by 
\[
t_P (t)  = \inf \{ s \ge t  \; | \; B (s-\Delta R) > X_{\rm off} \}\, . 
\]
Assuming a line rate of~$C$, transmissions can resume at 
\[
t_R (t)  = t_P (t) + \frac{X_{\rm off} -X_{\rm on}}{C} \, . 
\]

\paragraph{Round-trip Time (RTT)}
The round-trip time (RTT) refers to the total time it takes for a packet to travel from a source to a destination and back again.
Some congestion control algorithms such as TCP Vegas~\cite{Vegas} and Timely~\cite{timely} use measurements of the RTT to infer  whether congestion is growing or receding. 
In the path server model, the RTT is represented by the variable one-way delay from the source to the destination  plus the constant delay $\Delta R$. 
The one-way delay through the path server is represented by the horizontal distance between the departure function $D$ and the admitted arrival function $\hat{A}$. For the RTT measured at time $t$, denoted by ${\rm RTT} (t)$, we account for 
the delay $\Delta R$ by considering the horizontal distance at time $t - \Delta R$. 
Using  pseudo-inverse functions,  ${\rm RTT} (t)$ can be expressed 
as 
\begin{align}   
\begin{split}
{\rm RTT} (t) & = \bigl( ( t - \Delta R) - \hat{A}^{\downarrow}(D(t - \Delta R) \bigr) + \Delta R \\
& = t  - \hat{A}^{\downarrow}(D(t - \Delta R)) \, . 
\end{split}
\label{eq:RTT}
\end{align}

\subsection{Rate-Based Congestion Control}
\label{subsec:rate}

We next consider a rate-based CCA that sets the maximum transmission rate of the source. Since the rate is piecewise constant over  two congestion events, we can describe  the CCA rate limit  by an exact service curve of $S_{r} = \max(rt,0)$, where the  rate~$r$ changes according to congestion signals. 
Assuming that the rate is initialized to $r_o$, we therefore obtain 
\begin{align} 
\hat{A}(t)  = A \otimes S_{r_o} (t)
\quad \text{ and } \quad
D(t)  = \hat{A} \otimes S (t) \, .
\label{eq:conv-init-rate}
\end{align}


We consider a CCA that applies an Additive-Increase-Multiplicative-Decrease (AIMD) algorithm~\cite{Jain89} and  uses a timeout  as an indicator of congestion. Recall that a timeout event occurs when the acknowledgement function $D(t-\Delta R)$ falls below the timeout function $\hat{A} (t - \tau_o)$. In this case the rate is decreased proportionally by  
\begin{align}    
r(t) = \beta \cdot \, r(t^-)\, , 
\label{eq:mult-decr}
\end{align}
where $0 < \beta < 1$ is a constant. Here we use the notation 
$r(t^-)$ to indicate the rate just before the timeout occurred. An additive increase of the rate is achieved by allowing a  rate increase only after a fixed time interval,  which we denote by  $\Delta \tau_{\rm ai}$.  A rate increase occurs only if 
there was no timeout and no congestion notification in the previous interval, 
that is, for all~$s \in [t, t - \Delta \tau_{\rm ai}]$, 
$D(s-\Delta R) \ge \hat{A} (s - \tau_o)$ and $s-\Delta R$ does not satisfy the conditions for 
sending a congestion notification as described in~\S\ref{subsec:events}. 
The rate is increased by an additive constant $\alpha > 0$ by 
\begin{align}    
r(t) =  r(t^-)+\alpha \, . 
\label{eq:add-incr}
\end{align}

When the CCA updates the rate limit $r(t)$, the service curve $S_r$ changes,  
and the convolution expressions in Eq.~\eqref{eq:conv-init-rate} are no longer valid. By resetting the coordinate system and revising  function~$\hat{A}$, we can obtain a new min-plus linear system and continue the computation 
of the departure function~$D$. 

\begin{figure}[t!]
    \centering
    \subfloat[Timeout event at $t_{\rm TO}$.]{
        \includegraphics[width=0.48\textwidth]{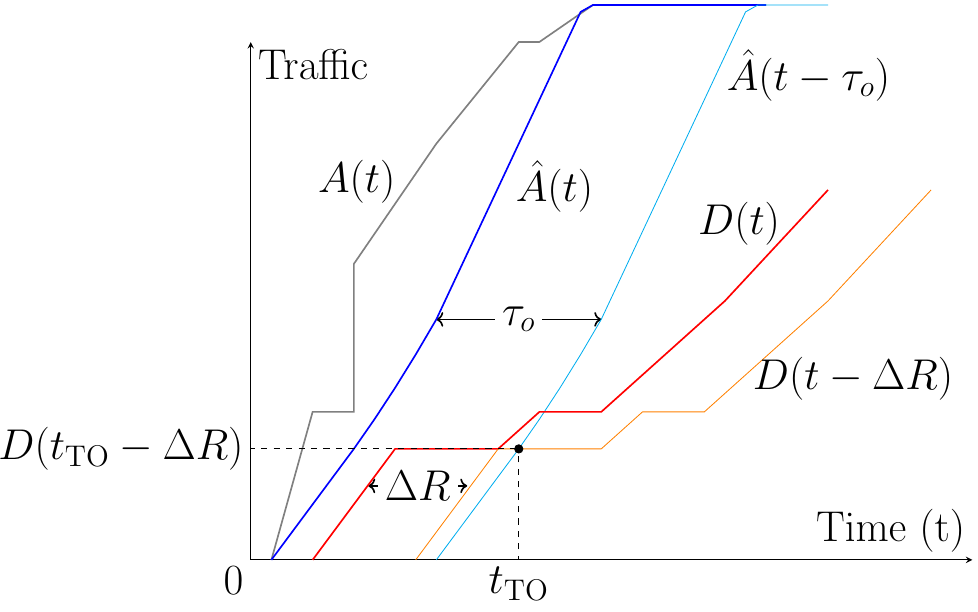}
        \label{fig:RateBasedTimeoutFound}
    }
    \subfloat[Coordinate shift to new origin at $(t_{\rm TO},D(t_{\rm TO}- \Delta R))$.]{
        \includegraphics[width=0.48\textwidth]{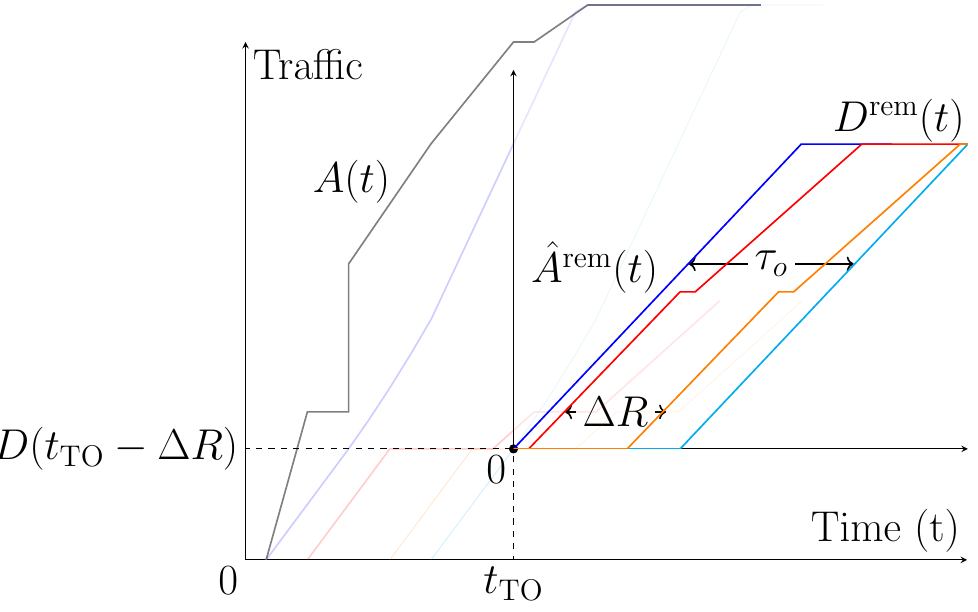}
        \label{fig:RateBasedTimeoutUpdate}
    }
     \caption{Updating the rate-based traffic functions after a timeout event. (Arrival  function~$A(t)$ in gray, admitted arrival function~$\hat{A}$  in blue, departure function~$D(t)$  in red, timeout function~$\hat{A}(t-\Delta R)$ in cyan, acknowledgement function~$D(t-\Delta R)$ in orange).}
     \label{fig:RateBasedTimeout}
\end{figure}

We  next illustrate the shift of the coordinate system for a timeout 
event and the subsequent retransmissions. 
Consider the scenario in Figure~\ref{fig:RateBasedTimeoutFound}, which 
has an arrival  function~$A(t)$ that is rate-limited to yield an admitted function~$\hat{A}$. The departures from the path server are described by~$D(t)$. At time $t_{\rm TO}$, the timeout function~$\hat{A}(t-\Delta R)$ becomes larger than the acknowledgement function~$D(t-\Delta R)$, indicating a timeout event. 
At this time, the CCA updates the rate limit according to Eq.~\eqref{eq:mult-decr}. 

The sender knows at time~$t_{\rm TO}$ that an amount given by 
$D(t_{\rm TO} - \Delta R)$  has been successfully delivered. Assuming a Go-Back-N scheme, the remaining unacknowledged packets,~$\hat{A}(t_{\rm TO})-D(t_{\rm TO}-\Delta R)$, need to be retransmitted. As a result, $\hat{A}$ is no longer a non-decreasing function and the min-plus convolution operation is no longer applicable. Additionally, after the timeout event, the rate limit is reduced according to  Eq.~\eqref{eq:conv-init-rate} and the service curve of the CCA becomes $S_{r(t_{TO})} = \max \{r(t_{TO}) t , 0\}$.

The retransmission as well as the multiplicative decrease of the rate limit can be captured through a change of the coordinate system, where we move the origin  to 
coordinates~$(t_{\rm TO}, D(t_{\rm TO}-\Delta R))$. This is shown in 
Figure~\ref{fig:RateBasedTimeoutUpdate}. 
In the new coordinate system, the traffic functions are given the superscript 
`rem' to indicate that they represent the remaining traffic. 
We set $A^{\rm rem}(t) = \hat{A}^{\rm rem}(t) = D^{\rm rem}(t) =0$ for $t \le 0$. 
The new arrival function $A^{\rm rem}$ corresponds to the traffic that has not been transmitted or been not acknowledged by time $t_{TO}$, yielding 
\begin{align*}
A^{\rm rem}(t) & = A(t+t_{\rm TO})-D(t_{\rm TO}-\Delta R)\, . 
\end{align*}
Note that the function has an initial burst $A(t)-D(t_{\rm TO}-\Delta R))$. 
Since the modified arrival function is non-decreasing we can apply 
the min-plus convolutions to obtain the admitted arrival and departure functions by 
\begin{align*}
\hat{A}^{\rm rem}(t)& = A^{\rm rem}\otimes S_{r(t_{\rm TO})} (t)\, , \\
D^{\rm rem}(t) & = \hat{A}^{\rm rem} \otimes S (t)\, .
\end{align*}
These functions can be used to update the original functions for times 
$t\geq t_{\rm TO}$ by 
\begin{align*}
\hat{A}(t)& = \hat{A}^{\rm rem}(t-t_{\rm TO})+D(t_{\rm TO}-\Delta R)\, , \\
D(t) & = D^{\rm rem}(t-t_{\rm TO}) +D(t_{\rm TO}-\Delta R) \, .
\end{align*}

A similar coordinate shift and update is performed whenever the 
rate limit is increased. For each time $t_{AI}$ where an additive increase according to~Eq.~\eqref{eq:add-incr} occurs we reset the origin to $(t_{AI}, \hat{A}(t_{AI}))$, which indicates the amount of traffic that has been admitted to the network so far.
Here, the modified arrival function $A^{\rm rem}$ becomes 
\begin{align*}
A^{\rm rem}(t)  = A(t+t_{\rm AI})-\hat{A} (t_{\rm AI}) \, .  
\end{align*}
After updating the rate limit to $r (t_{\rm AI})$, the admitted arrivals and departures are computed with the min-plus convolution as shown above. 
The computations for the rate-based congestion control are summarized in an algorithm in Appendix~\ref{sec:rate-algorithm}.

\subsection{Window-Based Congestion Control}
\label{subsec:window}

With  window-based congestion control, the CCA limits the total number of bytes 
that can have outstanding acknowledgements, the so-called congestion window.  The  majority of CCAs for TCP, from  Tahoe~\cite{Tahoe} to CUBIC~\cite{Cubic},  follow a window-based approach. 
A major difference between rate-based and window-based CCAs is that the latter 
allows back-to-back transmissions, which increases the burstiness of traffic. Here, we model a basic window-based CCA with AIMD, similar to Reno~\cite{Floyd_1996} for the path server model from Figure~\ref{fig:NetworkCalculusModel}. 
 
Let $W$ denote the size of the congestion window, the exact service curve associated with window-based congestion control is given by 
\[
S_W (t) = 
\begin{cases}
    0 \, , & \text{ if } t \le 0 \, , \\
    W \, , & \text{ if } t > 0 \, . \\
\end{cases}
\]
With this service curve, the admitted arrival  and departure functions are computed as 
\begin{align} 
\hat{A}(t)  = A \otimes S_{W} (t)
\quad \text{ and } \quad
D(t)  = \hat{A} \otimes S (t) \, .
\label{eq:conv-init-window}
\end{align}

In practice, window-based CCAs update $W$ each time an acknowledgement of previously unacknowledged data is received. In our characterization, we update the congestion window only after~$W$ bytes have been acknowledged. The amount of data corresponding to~$W$ is referred to as a {\it flight}.
Starting with an initial congestion window~$W_o$, the transmission of the initial flight is completed at~$t_W$ given by  
\[
t_W = \inf \left\{ t > 0 \; \mid \; D(t- \Delta R) \ge W_o \right\} \, . 
\]
With an additive increase, after the transmission of a flight, the congestion window is 
increased by a constant $\alpha>0$, that is, 
\begin{align}
    W =  W_o+ \alpha\, . 
    \label{eq:add-incr-window}
\end{align}
After a  timeout, that is, a 
time~$t_{\rm TO}$ with $D\left(t_{\rm TO}-\Delta R\right)<\hat{A}(t_{\rm TO} - \tau_o)$, the congestion window is  reduced by a multiplicative 
factor $\beta$ ($0<\beta<1$), that is, $W = \beta \, W_o$. 

Whenever the congestion window is updated according to Eq.~\eqref{eq:add-incr-window},  we perform a shift of the coordinate system. 
After the transmission of a flight, the origin of the coordinate system 
is set to~$(t_W, \hat{A}(t_W))$. 
After a timeout at some time $t_{\rm TO}$, the coordinate system is adjusted as described in \S\ref{subsec:rate} with a new origin~$(t_{\rm TO}, D(t_{\rm TO}-\Delta R))$. 
In both cases, due to resetting the origin, the updated congestion window becomes the initial window size, that is, $W_o = W$. 
The remaining traffic functions $A_{\rm rem}$, $\hat{A}_{\rm rem}$, and $D_{\rm rem}$ in the new coordinate system are constructed as described in \S\ref{subsec:rate}. 

Some CCAs use a multiplicative increase when the 
window size is below a value $w_{\it th} < W$. In TCP Tahoe, Reno, and its variants, the  value is referred to as slow start threshold, and a time period of a multiplicative increase is referred to as a slow start phase.  
Typically, $W$ is doubled after a flight, resulting in an update  
\begin{equation}
W  = 
\begin{cases}
    2 \, W \, , & \text{ if } W < w_{\it th}   \, , \\
    W +a  \, , & \text{ if } W \ge w_{\it th}  \, . \\
\end{cases}
\label{windowFunction}
\end{equation}
After a timeout event, the slow start threshold is reduced by a multiplicative factor
($w_{\it th} = \beta W$) and the congestion window  is reduced to one packet with maximal length. Appendix~\ref{sec:window-algorithm}  presents an algorithm for the network calculus computations for window-based congestion control.

\subsection{Example: TCP Vegas}

\begin{figure}[t!]
    \subfloat[Simulation.]{
        \includegraphics[width=0.48\textwidth]
        {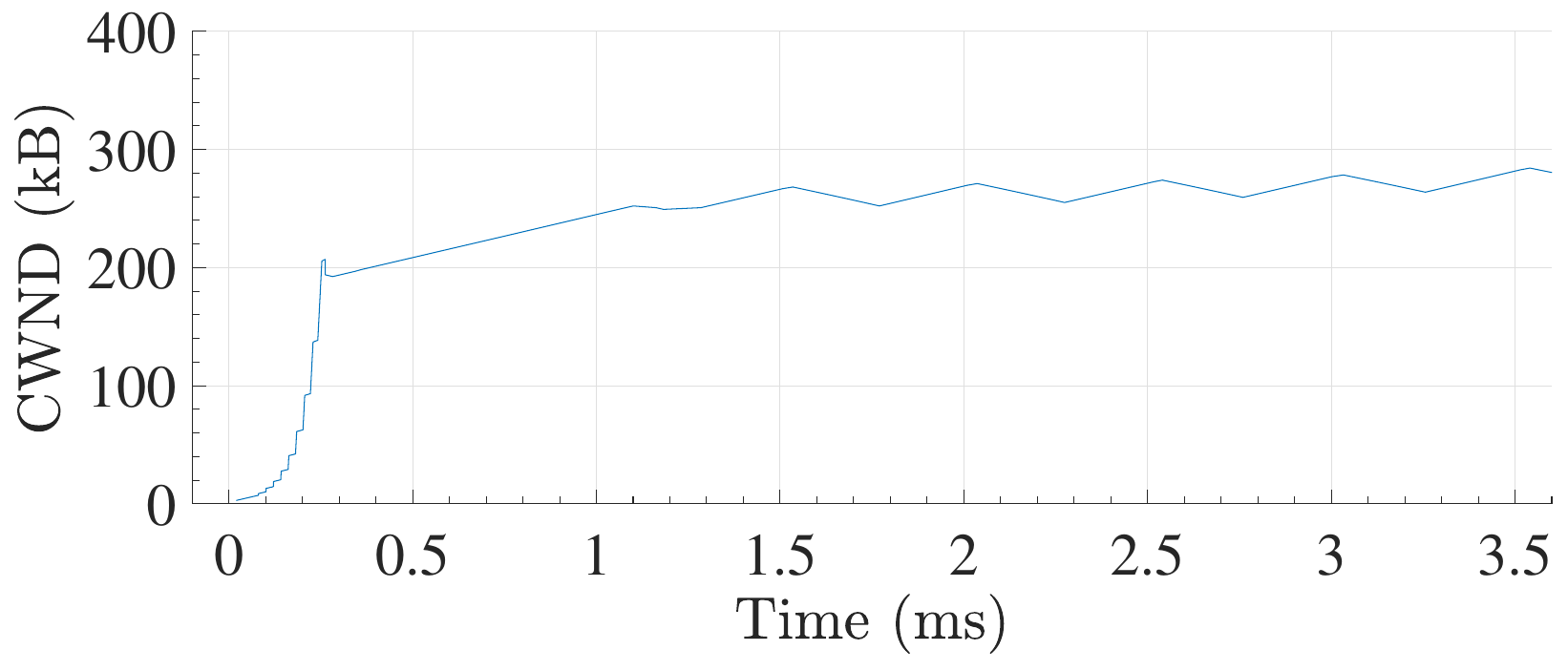}
        \label{fig:NS3VegasCWND}
    }
    \centering
    \subfloat[Model-based.]{
        \includegraphics[width=0.48\textwidth]{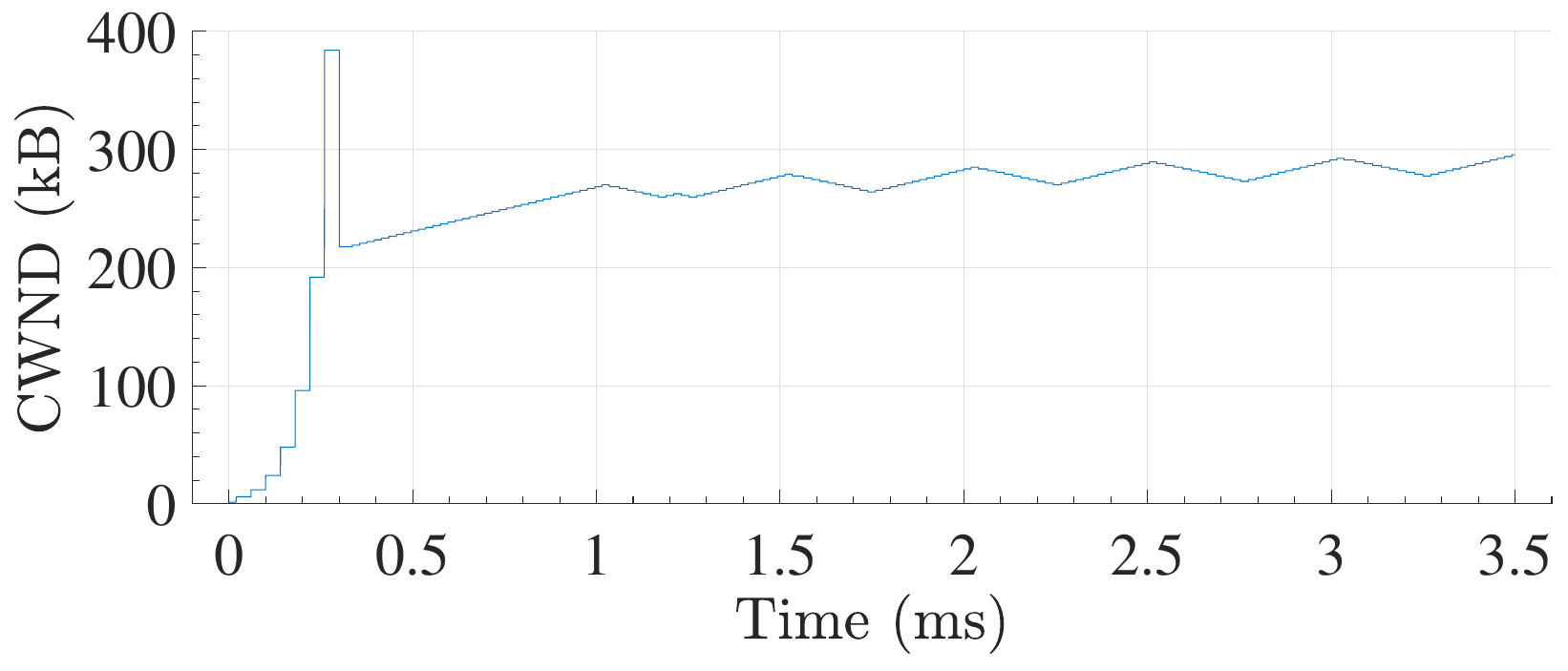}
        \label{fig:NCVegasCWND}
    }

     \caption{Congestion window (CWND) of TCP Vegas.}
     \label{fig:VegasCWND}
\end{figure}

\begin{figure}[t!]
    \subfloat[Simulation.]{
        \includegraphics[width=0.48\textwidth]
        {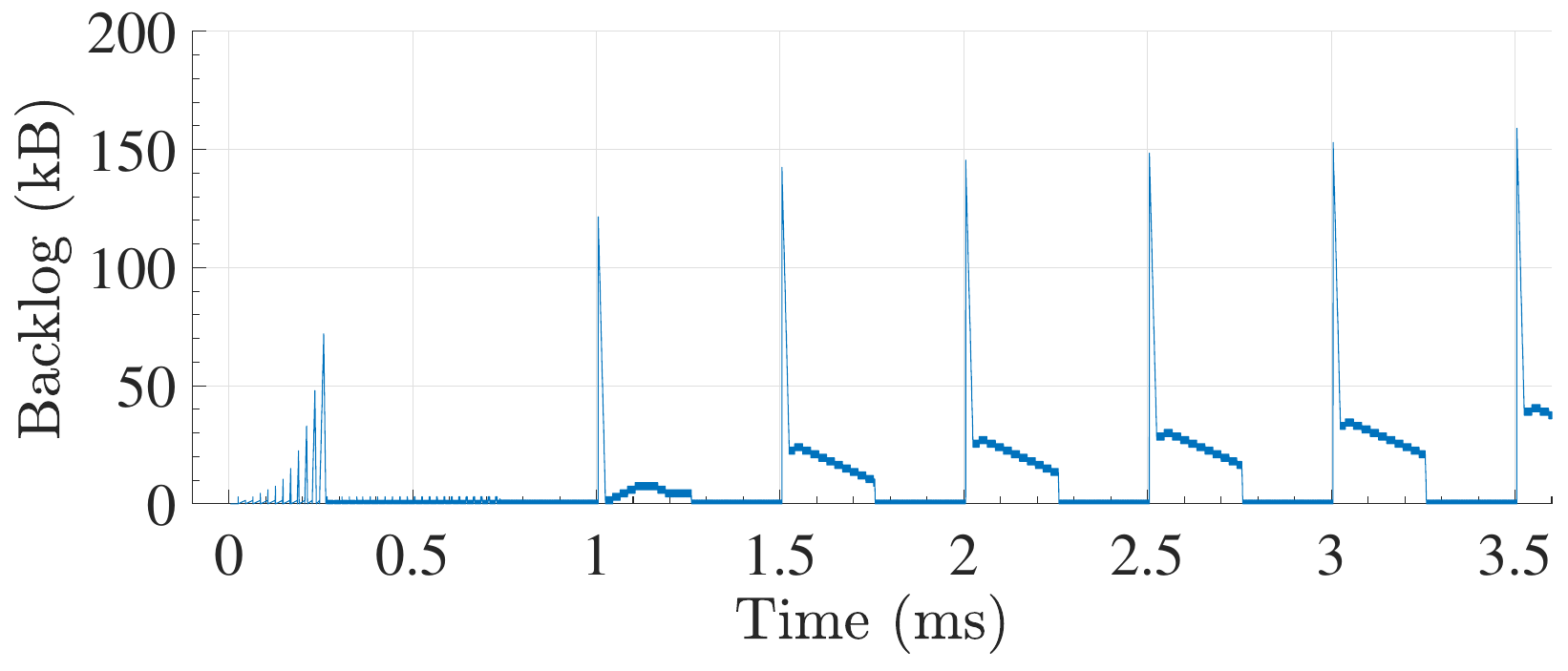}
        \label{fig:NS3VegasBacklog}
    }
    \centering
    \subfloat[Model-based.]{
        \includegraphics[width=0.48\textwidth]{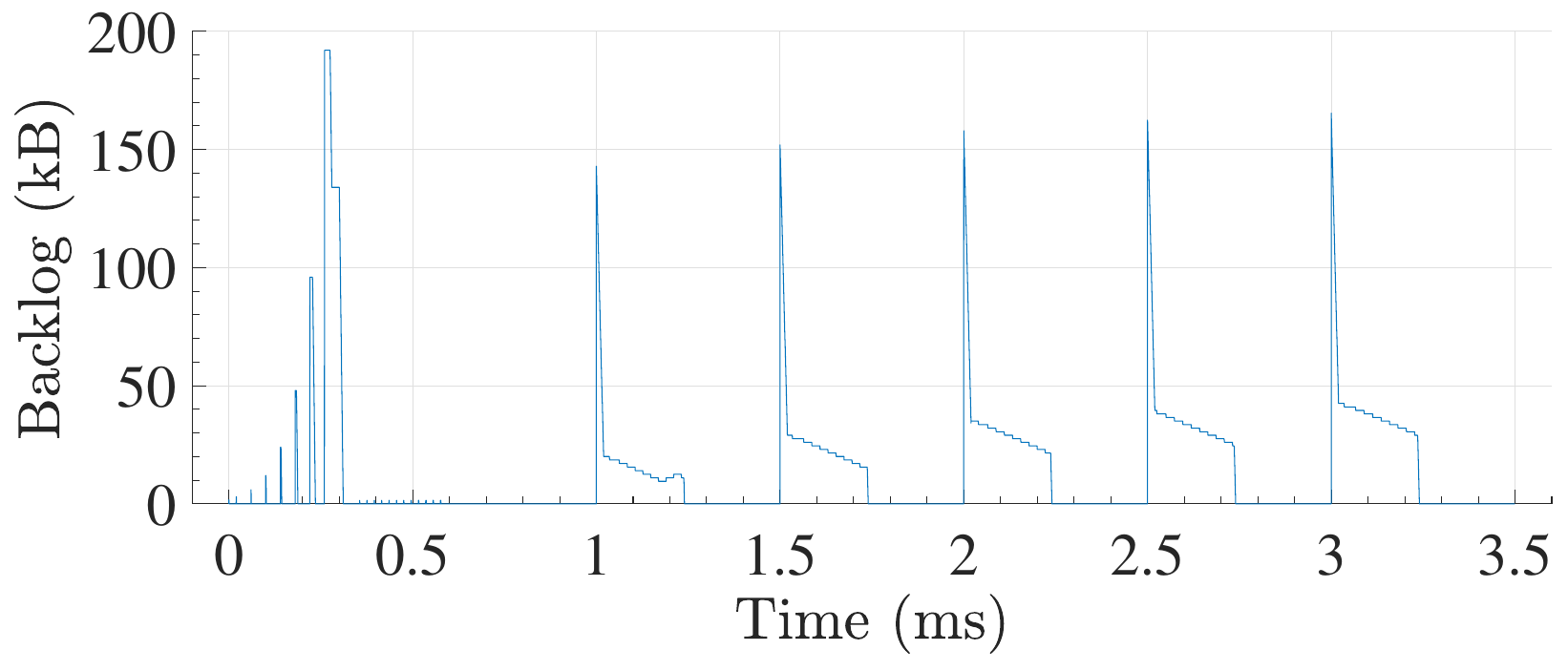}
        \label{fig:NCVegasBacklog}
    }
     \caption{Backlog produced by TCP Vegas with our traffic scenario.}
     \label{fig:VegasBacklog}
\end{figure}

To evaluate how well CCA dynamics are captured by our network calculus model, 
we compare  a model-based characterization of TCP Vegas~\cite{Vegas} with a packet-level simulation. TCP Vegas is a window-based CCA that uses RTT estimates for  updating its congestion window with the goal of keeping the backlog within a specified range. It has a slow-start phase, which doubles the congestion window every other round-trip time. 

We consider a bursty transmission scenario of a single traffic source at a 100~Gbps link, which initially issues a 4~MB burst and then transmits at fixed rate of 50~Gbps. Additionally, starting at $t=1~ms$, the source issues instantaneous bursts of 1.5~MB every 0.5~ms. Each of the bursts creates a backlog at the link. 

For the simulations, we use the TCP Vegas implementation included in the ns-3 distribution~\cite{ns3} and set protocol parameters as in~\cite{vegasVernon}. 
The network topology consists of the source, a switch, and a destination node. 
Links have a propagation delay of~$5~\mu s$, resulting in a round-trip delay between the source and the destination of~$20~\mu s$.
Figures~\ref{fig:NS3VegasCWND} and~\ref{fig:NS3VegasBacklog}, respectively, present the 
congestion window at the source and the backlog at the switch. 

Our model-based characterization uses a path server with 
fixed rate of \, $C=100$~Gbps and feedback delay $\Delta R=20~\mu$s. 
The computations in \S\ref{subsec:window} are modified to account for RTT measurements and different congestion window updates.  Appendix~\ref{sec:vegas-algorithm} presents an algorithm of the calculations for TCP Vegas. Figures~\ref{fig:NCVegasCWND} and~\ref{fig:NCVegasBacklog} depict the congestion window and the backlog 
computed by the model. 

There are two deviations between the simulation and the network calculus model. 
The first difference concerns the {\it baseRTT} parameter. The simulation sets it to the smallest RTT measurement since the connection started, whereas the model sets it to the feedback delay $\Delta R$. The second difference relates 
to the window increases during the slow start phase. The ns-3 simulation updates the window for each new acknowledgement, whereas our model follows the TCP Vegas specification in~\cite{vegasVernon,Vegas} and doubles the window for every two RTT updates.

A comparison of Figures~\ref{fig:VegasCWND} and~\ref{fig:VegasBacklog} shows  that the model-based characterization closely matches the simulation. 
The largest deviation, observed in the interval $[0,0.5]$~ms, is due to the different window updates during the slow start phase. 
Beyond this, the model captures the variations and trends of the congestion window during  burst arrivals.

\section{Multi-flow Analysis}
\label{sec:multiflow}

\begin{figure}[t]
    \centering
    \includegraphics[width=0.9\textwidth]{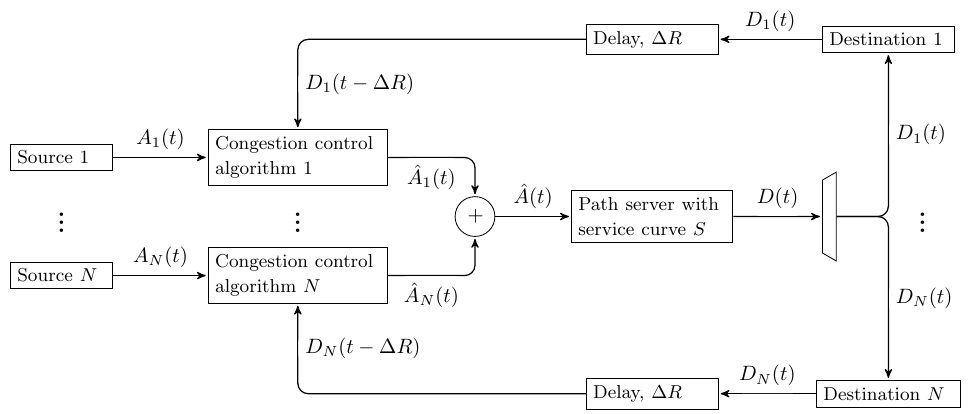}
    \caption{Multi-flow model for congestion control.}\label{fig:MultipleFlowNetworkCalculusModel}
\end{figure}

To study fairness properties of CCAs we must extend the 
path server model from Figure~\ref{fig:NetworkCalculusModel}  to accommodate multiple flows. The extended model is shown in Figure~\ref{fig:MultipleFlowNetworkCalculusModel}, 
where multiple traffic sources that are each regulated by a CCA 
are multiplexed at the path server resulting in the admitted arrival function~$\hat{A} (t) = \sum_{j=1}^N \hat{A}_j (t)$. 
At the egress of the path server, the departure traffic $D$ is 
demultiplexed into per-flow departure functions~$D_j$, 
which are then used for the acknowledgement functions. 
For simplicity, we assume that the  feedback delay $\Delta R$ is identical for all flows. 

Obtaining  expressions for $D_j$ is challenging, since it depends on the operation of the path server, in particular, how the path server schedules traffic from different flows. In the next subsection, we present a derivation of $D_j$ for a path server with FIFO multiplexing. Obtaining expressions for $D_j$ under priority or fair queuing involves less effort. 
 
\subsection{Demultiplexing Departures at a FIFO Element}
\label{subsec:dmux-fifo}

Consider a path server as shown in Figure~\ref{fig:MultipleFlowNetworkCalculusModel}, which sees arrivals from a set ${\mathcal N} = \{1, \ldots, N \}$ of flows, and which transmits traffic in FIFO order.  
Since, with FIFO, traffic departs in the order of its arrival,  
if at some time $t$ the traffic of a flow that arrived before time $t' \le t$ departs, it also holds for every other flow.
We call this the {\it FIFO property}. 

\begin{definition} {\bf (FIFO property) }
	A path server that sees arrivals from a  set of flows~${\mathcal N}$ has the \emph{FIFO property} if, for every~$i,j\in {\mathcal N}$ and every time interval~$[t', t]$, 
	\begin{equation}
		\label{eq:fifo_cond}
		\hat{A}_i(t') < D_i(t) \implies \hat{A}_j(t') \le D_j(t)\,.
	\end{equation}
\end{definition}
The strict inequality on the left-hand side of Eq.~\eqref{eq:fifo_cond} is needed to account for simultaneous packet arrivals. Note that in FIFO, the order of departures between simultaneous arrivals is arbitrary. 
If the arrival function is continuous or if simultaneous packet arrivals from multiple flows can be excluded, the strict inequality can be relaxed to allow for equality.

The following lemma shows that the FIFO property can be extended to an aggregation of flows. The proof of the lemma as well as other proofs in this section are 
given in Appendix~\ref{sec:proofs}. 
\begin{lemma}
	\label{lemma:aggregate_fifo_property}
	Given a path server with~${\mathcal N}$ flows that satisfies the FIFO property.
    For any subset~$X\subseteq {\mathcal N}$ and any flow~$i\in {\mathcal N}$, every time interval~$[t', t]$ satisfies 
	\begin{align}
        \label{eq:aggregate_fifo_property:part_imply_sum}
        \hat{A}_i(t') < D_i(t) &\implies \sum_{j\in X} \hat{A}_j(t') \le \sum_{j\in X} D_j(t)\,, \text{ and }\\
        \label{eq:aggregate_fifo_property:sum_imply_part}
		\sum_{j\in X} \hat{A}_j(t') < \sum_{j\in X} D_j(t) &\implies \hat{A}_i(t')\le D_i(t)\,.
	\end{align}
\end{lemma}
Similar to Eq.~\eqref{eq:fifo_cond}, if the arrival function is continuous, or if or there are no simultaneous packet arrivals from different flows, the left equations of Eq.~\eqref{eq:aggregate_fifo_property:part_imply_sum} and~\eqref{eq:aggregate_fifo_property:sum_imply_part} can be relaxed to allow for equality. Then, the implications in the lemma can be combined into
\begin{equation*}
     \hat{A}_i(t') \le D_i(t) \iff \sum_{j\in X} \hat{A}_j(t') \le \sum_{j\in X} D_j(t)\,.
\end{equation*}
If the aggregate arrival and departure functions~$\sum_{j\in X} \hat{A}_j(t')$ and~${\sum_{j\in X}D_j(t)}$ coincide for  two times~$t'$ and $t$ with $t' < t$, we can calculate the departure function of individual flows. 
\begin{lemma}
    \label{lemma:strict_aggregate_fifo_property}
    Given a buffer with~${\mathcal N}$ flows that satisfies the FIFO property.
    For any subset~$X\subseteq {\mathcal N}$ and every time interval~$[t', t]$, if
	\begin{equation}
        \label{eq:strict_aggregate_fifo_property:antecedent}
		\sum_{j\in X} \hat{A}_j(t') = \sum_{j\in X} D_j(t)\,,
	\end{equation}
    then for any flow~$i\in {\mathcal X}$,
    \begin{equation}
        \hat{A}_i(t') = D_i(t)\,.
    \end{equation}
\end{lemma}

Using Lemmas~\ref{lemma:aggregate_fifo_property} and~\ref{lemma:strict_aggregate_fifo_property}, we can calculate the bounds of the departure function of individual flows from a FIFO buffer.
\begin{theorem}
	\label{thm:split_fifo}
	Let~${\mathcal N}$ be a set of flows at a buffer that satisfies the FIFO property.  
 Then the departure function of every flow~$i\in {\mathcal N}$ is bounded for 
 arbitrary~$\varepsilon>0$ by 
	\begin{equation*}
		\hat{A}_i(\hat{A}^{\uparrow}(D(t))) \le D_i(t) \le \hat{A}_i(\hat{A}^{\uparrow}(D(t)) + \varepsilon)\,.
	\end{equation*}
\end{theorem}
If~$\hat{A_i}$ is continuous, the upper bound in Theorem~\ref{thm:split_fifo} reduces to the lower bound for~$\varepsilon \to 0$, and the value of~$D_i(t)$ can be computed exactly.
\begin{figure}[t]
    \centering
        \subfloat[Transmission rates.]  {\includegraphics[width=0.35\textwidth]{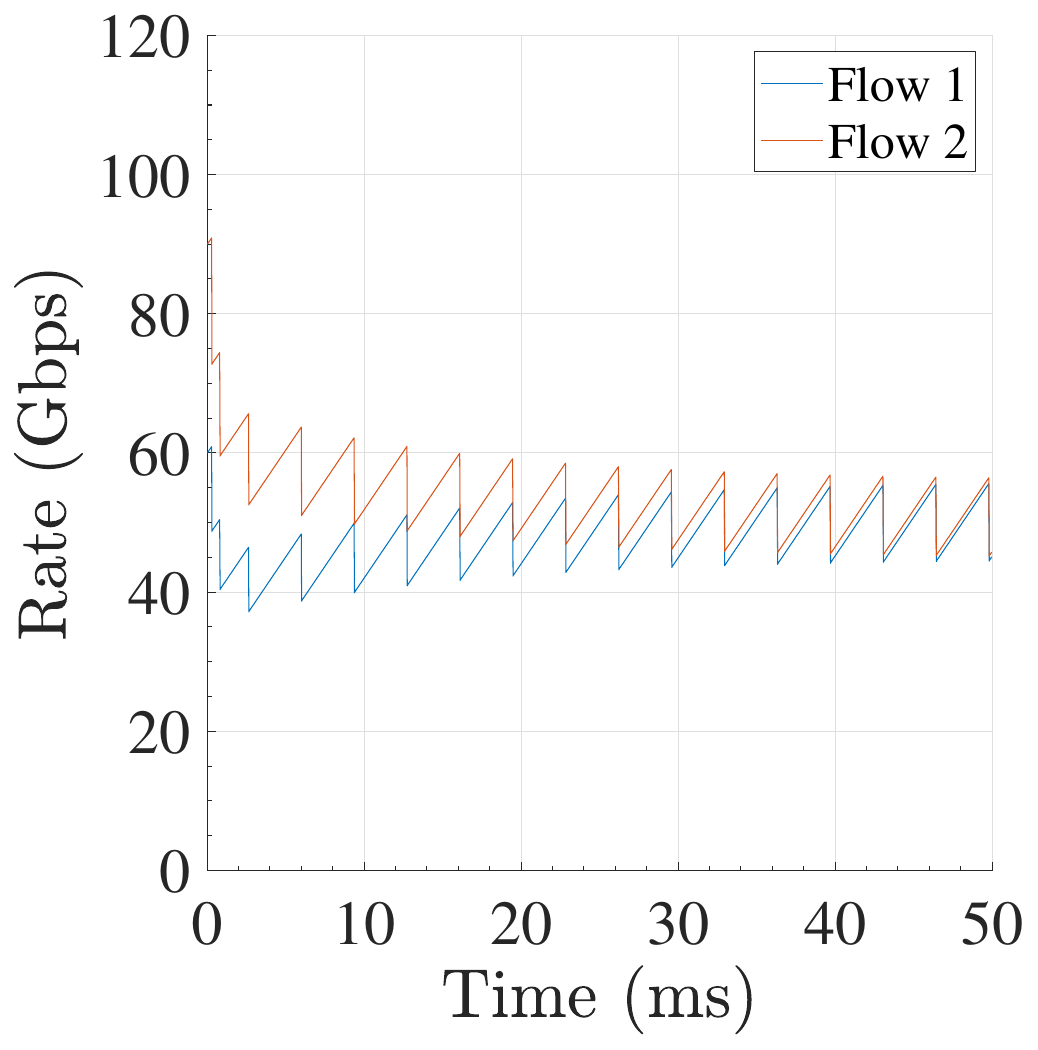}
     \label{fig:AIMDfairnessRates}
    }
    \hspace{1cm}
    \subfloat[Convergence to fairness and efficiency.]{\includegraphics[width=0.35\textwidth]{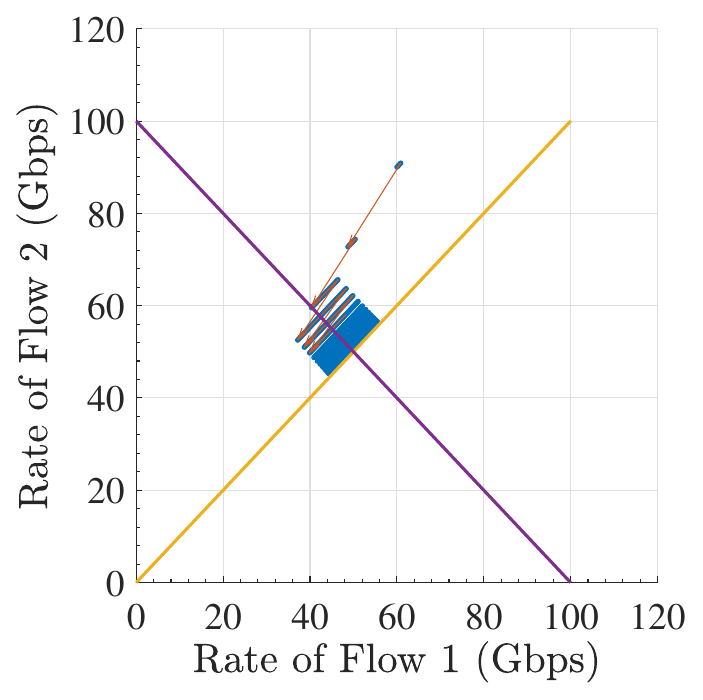}
    \label{fig:AIMDfairness}}

    \caption{Fairness at a FIFO path server.}
\end{figure} 
\subsection{Example: AIMD Fairness}

With~\S\ref{subsec:dmux-fifo} 
we can analyze the dynamics of congestion control for 
concurrent flows. We now present an example that evaluates how well the model-based network calculus characterization  tracks a convergence 
of a CCA to a fair allocation. 
We consider a scenario with two flows that each deploy a  rate-based AIMD algorithm as in \S\ref{subsec:rate}
with $\alpha = 100$~Mbps and $\beta = 0.8$. The time interval between rate increases is~$\Delta \tau_{ai} = 30~\mu$s. 
The feedback delay is set to $\Delta R = 20$~$\mu$s and the timeout value is $\tau_o = 100$~$\mu$s. 
The path server is assumed to have an exact service curve~$S(t) = \max\{Ct, 0\}$ with $C=100$~Gbps. There are two flows that start transmissions with rates~60~Gbps and 90~Gbps.

The computations apply Algorithm~\ref{alg:AIMD-departures} based on the derivations in \S\ref{subsec:rate}. Figure~\ref{fig:AIMDfairnessRates} shows the rates of the two flows over a time period of 50~ms. 
Figure~\ref{fig:AIMDfairness} depicts a vector representation of the rates  as suggested in~\cite{Jain89}. The figure includes two lines: a fairness line (yellow) where both flows are allocated the same rate, and an efficiency line (purple) where both flows consume all link resources. The intersection of the two line is the optimal operating point. Starting at the initial point 
$(60,90)$, the shown graph shows the convergence to the optimal operating point using a quiver plot. 
Clearly,  the model-based characterization is able to capture 
the convergence to a fair resource allocation.

\section{Case Study: Data Center Congestion Control}
\label{sec:casestudy}

In this section we consider a congestion control scenario in a data center network, where we  compare a packet-level simulation with our network calculus analysis. 
We consider a data center network with RDMA transport over RoCEv2, which performs RDMA with UDP encapsulation~\cite{rocev2}.  RoCEv2 networks support a lossless Ethernet service which is realized with the PFC protocol (see \S\ref{subsec:events}). 
The de-facto standard for congestion control in these networks is 
Data Center Quantized Congestion Notification (DCQCN)~\cite{DCQCN-Sigcomm2015}, 
which is supported by numerous vendors~\cite{dcqcn-nvidia,dcqcn-cisco,dcqcn-juniper,dcqcn-huawei}.  Unless PFC is disabled, it runs in conjunction with DCQCN. 

At its core, DCQCN is a rate-based CCA with  ECN and RED for congestion notifications. 
DCQCN overall follows an AIMD approach for setting the transmission rates of senders. Congestion notification packets (CNPs) sent to senders must be separated by a gap of at least $T_{\rm gap}$. In detail, the rate adjustment at the senders is as follows:
 \begin{itemize}
		\item A sender maintains two rate limits~$R_C$ and~$R_T$. Upon receiving a~CNP the limits are updated by 
  $R_T = R_C$, 
  $R_C     = (1 - \alpha)R_C$, and 
$\alpha  = \alpha + g(1 - \alpha)$, where $\alpha \in [0, 1]$ is a rate reduction factor, which itself is tuned by a  parameter $g\in [0, 1]$.  Initially, $R_T$ and $R_C$ are set to the line rates, and $\alpha$ is set to~1.

		\item When a time~$K$ has elapsed since  the last CNP arrival, the sender sets $\alpha = (1 - g)\alpha$. 

		\item Every~$T$ time units without receiving a~CNP, a counter $i_T$ is increased by~1, and 
for every~$B$~bytes  transmitted without receiving a CNP, $i_B$ is increased by 1.  After~5 increments of~$i_T$ or~$i_B$, the sender updates the transmission rates $R_C$ and $R_T$ using the additive constants $R_{AI}$ and $R_{HI}$.  The rate increases of  $R_C$ and $R_T$ depend  on the current values of $i_T$ or~$i_B$ and are as given in Table~\ref{table:aimd}. Whenever a sender receives a CNP, it sets $i_B=i_T=0$.
	\end{itemize}
Table~\ref{table:PFC-DCQCN-parameters} lists parameter values for DCQCN  for a DCN switch as suggested in~\cite{DCQCN-Sigcomm2015,qcn}.
For PFC we set $X_{\rm off}=	9.5$~kB/port/Gbps and 
$X_{\rm on}= 	9.25$~kB/port/Gbps. 

 \begin{table}[!t]
     \caption{Increase of transmission rates in DCQCN.}
	\begin{center}
	\begin{tabular}{  ccl  }
      \hline  
        Phase & Condition & Changes \\
        \hline 
		{ Fast recovery}&$\max \{ i_T, i_B\} < 5$ & $R_T$  unchanged, $R_C = (R_C + R_T)/2$ \\[3pt]
        { Additive increase} & $\max \{ i_T, i_B\} \ge 5$ & $R_T = R_T + R_{AI}$,  $R_C = (R_C + R_T)/2$ \\[3pt]
        { Hyper increase} & $\min \{ i_T, i_B\}  \ge 5$ & $R_T = R_T + (\min\{i_T, i_B\}-5)R_{HI}$ \\
		 && $R_C = (R_C + R_T)/2$ \\
  \hline
	\end{tabular}
	\end{center}
\label{table:aimd}
\vspace{15pt}

\caption{Parameter setting for DCQCN from~\cite{DCQCN-Sigcomm2015,qcn}.}
\begin{center}
\begin{tabular}{lllllllll}
\hline 
$K_{\rm min}$ & $K_{\rm max}$ & $P_{\rm max}$ & $g$  & $T_{\rm gap}$ & 
$K, T$ & $B$ & $R_{AI}$ & $R_{HI}$ \\
\hline
5 kB & 200 kB  & 	1\% &	1/256 & 50 $\mu$s & 55 $\mu$s & 10 MB & 5 Mbps & 50 Mbps\\
\hline 
\end{tabular}
\end{center}
\label{table:PFC-DCQCN-parameters}
\end{table}%


\paragraph{Transmission scenario.} 
We consider a transmission scenario that may occur during  
distributed training of deep neural networks, when 
multiple workers  concurrently transmit gradients of 
model parameters through the same output port of a 
switch. 
We consider a shared memory switch with 32~ports that operate at 100\,Gbps. 
Each port is connected to a host, of which 31 hosts 
are traffic senders and one host is the destination for all traffic. The  propagation delay of  links between hosts and the switch is set to 1\,$\mu$s. The switch memory is assumed to be large enough so  that buffer overflows do not occur. We consider a scenario where 
all traffic senders start 
the transmission of a burst with size 10\,MB, which  
corresponds to the parameter size of the largest convolution layer in  ResNet50~\cite{resnet50}. 
We consider three network configurations: 
\begin{center}
\begin{tabular}{l c l}
{\it PFC } & -- & PFC is enabled, DCQCN is not enabled, \\
{\it DCQCN} & -- & DCQCN and PFC are enabled, \\
{\it DCQCN-noPFC} & -- & DCQCN is enabled, PFC is not enabled. 
\end{tabular}
\end{center}

\begin{figure}[t!]
\centering 
\subfloat[Total arrival rate.]{\includegraphics[width=0.48\textwidth]{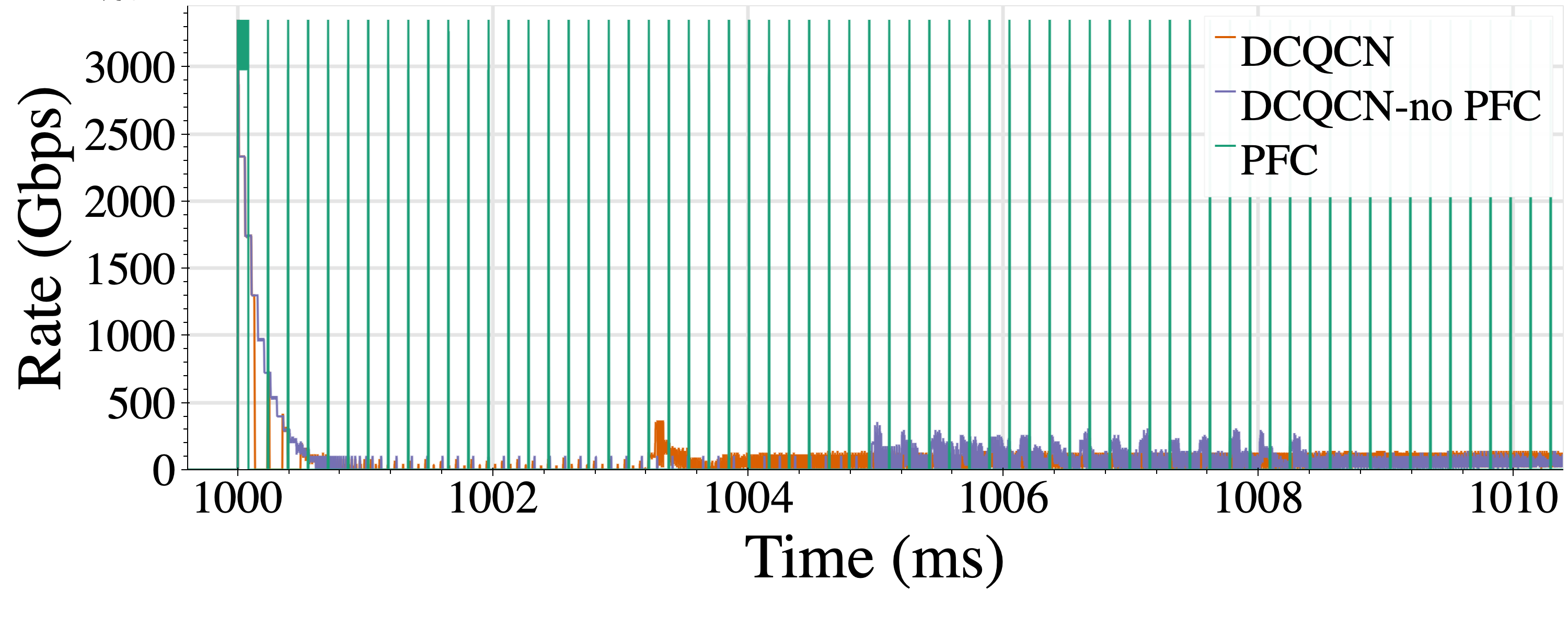}\label{fig:oneburst_31workers_delay1us_total10ms_rate}}
\centering 
\subfloat[Backlog.]{\includegraphics[width=0.48\textwidth]{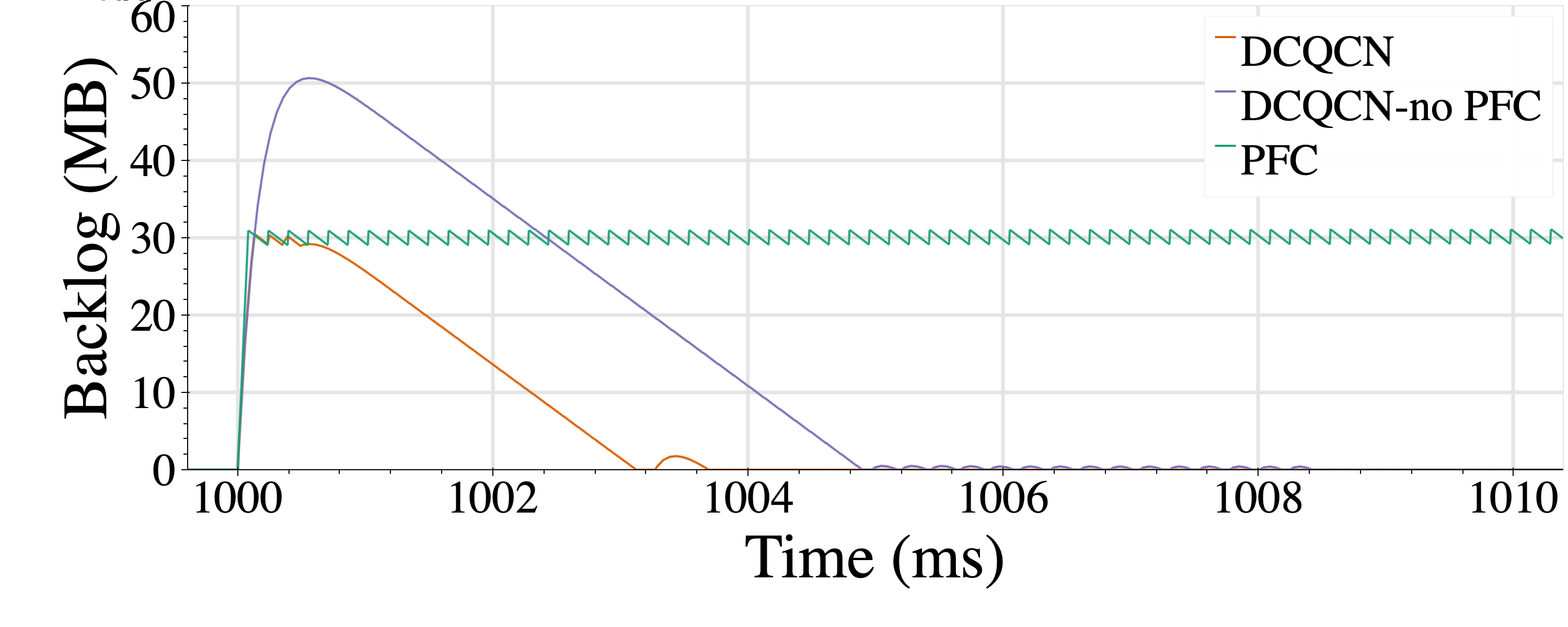}\label{fig:oneburst_31workers_delay1us_total10ms_backlog}}

\caption{Simulation of burst arrivals (10~ms).} \label{fig:oneburst_31workers_delay1us_total10ms}
\centering 
\subfloat[Total arrival rate.]{\includegraphics[width=0.48\textwidth]{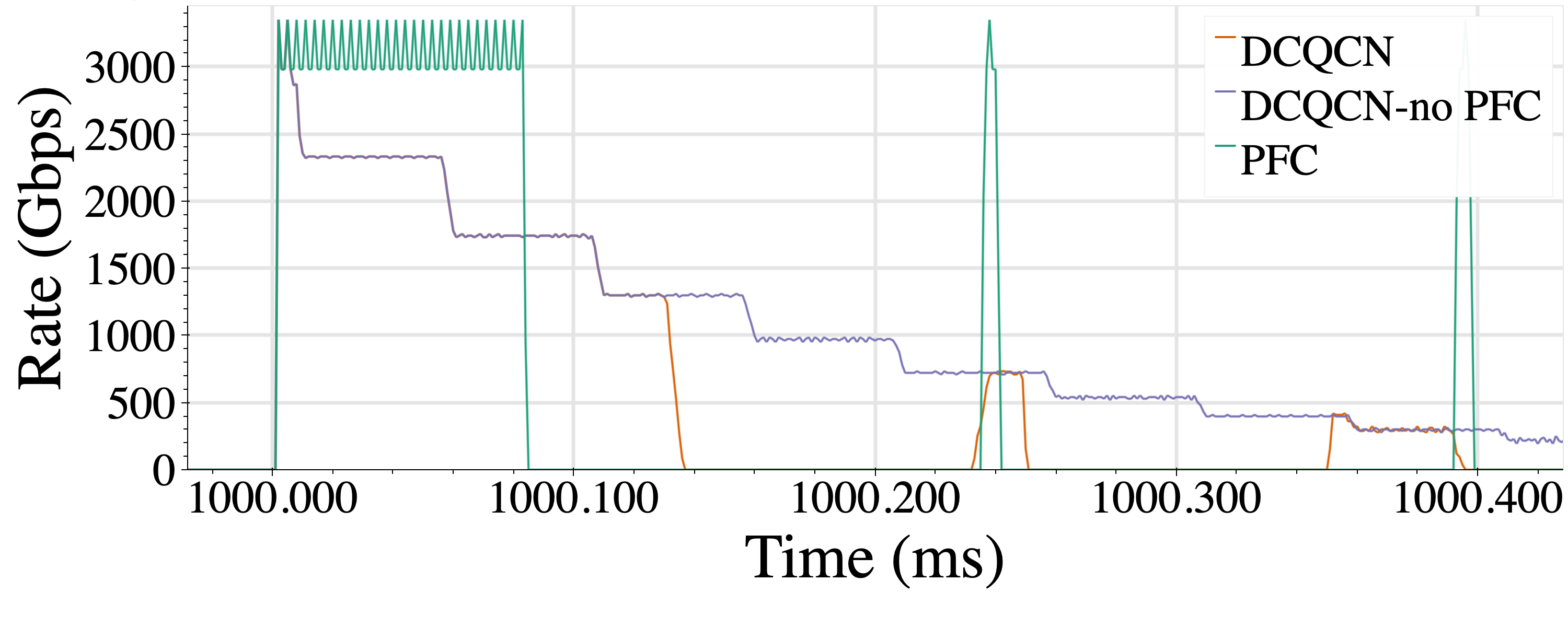}\label{fig:oneburst_31workers_delay1us_total400us_rate}}
\centering 
\subfloat[Backlog.]{\includegraphics[width=0.48\textwidth]{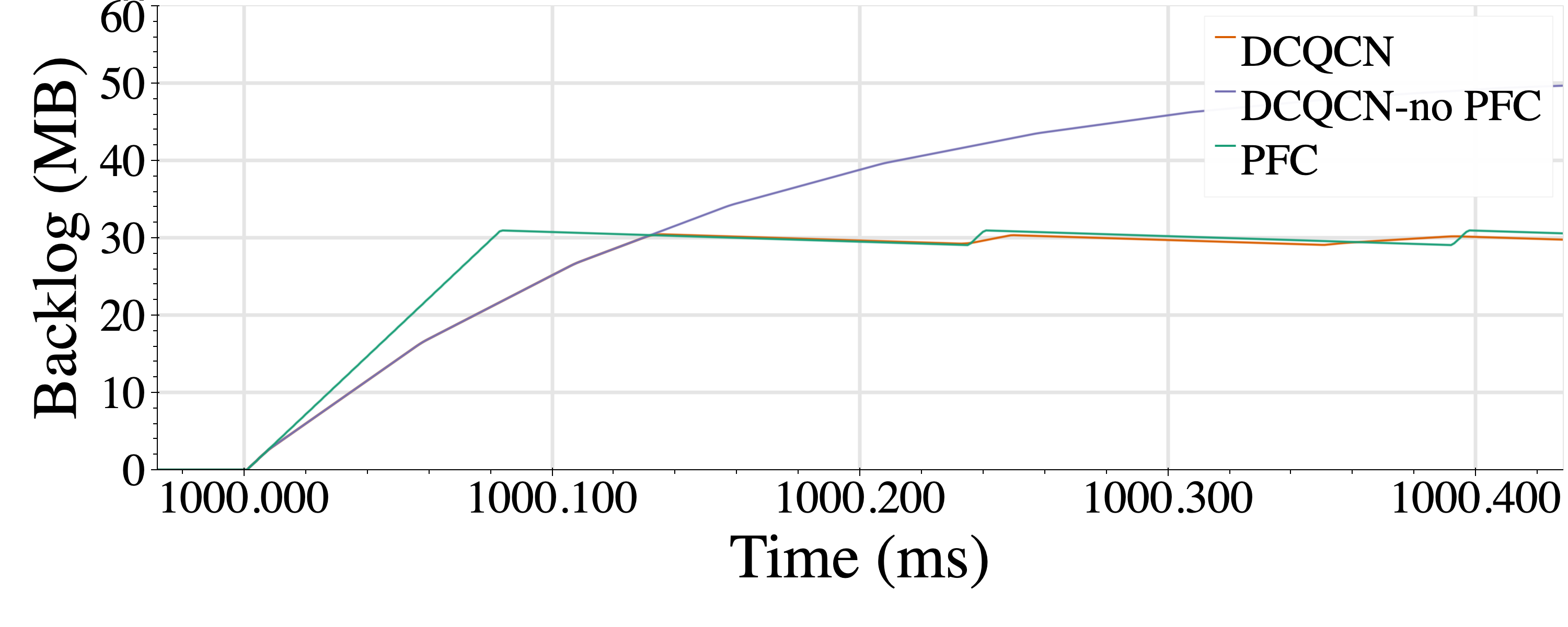}\label{fig:oneburst_31workers_delay1us_total400us_backlog}}

\caption{Simulation of burst arrivals (400~$\mu$s).} \label{fig:oneburst_31workers_delay1us_total400us}
\end{figure}

\subsection{Packet-level Simulation Results}
\label{subsec:dcqcn-sim}

We perform simulations using a publicly available 
ns-3 simulation for DCQCN and PFC over RoCEv2~\cite{rdmasim-code} . 
Figure~\ref{fig:oneburst_31workers_delay1us_total10ms_rate} shows the aggregate traffic arrival rate  at the switch for a duration of~10\,ms, starting at $t = 1000$\,ms, where 
rates are computed as averages over 1\,$\mu$s,  
 and Figure~\ref{fig:oneburst_31workers_delay1us_total10ms_backlog} 
shows the resulting backlog at the switch. 

The transmission rates  for {\it PFC} in Figure~\ref{fig:oneburst_31workers_delay1us_total10ms_rate} (shown in green color) exhibit an on-off behavior. Since the aggregate traffic to the port with the destination host exceeds its link rate, backlog accumulates at the switch. When the threshold $X_{\rm off}$ is reached, arrivals are stopped until the backlog falls below $X_{\rm on}$. Since all senders start transmissions at the same time, the pause and resume events are synchronized, which creates the observed periodic traffic pattern. 
The resulting backlog for {\it PFC} seen in  Figure~\ref{fig:oneburst_31workers_delay1us_total10ms_backlog} shows that the backlog initially builds up to around 30\,MB and then stays at a high level. 

The results for {\it DCQCN} and {\it DCQCN-noPFC}, respectively, are shown as blue and red curves in Figure~\ref{fig:oneburst_31workers_delay1us_total10ms}. 
Initially, the traffic rates are high for both configurations,  
since DCQCN sets the initial transmission rates to the line rate of 100\,Gbps.  Within one millisecond, by time $t = 1001$\,ms,  the total traffic rate is reduced significantly and remains low for the remaining duration. 

The evolution of the backlog for {\it DCQCN} in Figure~\ref{fig:oneburst_31workers_delay1us_total10ms_backlog} shows a large backlog for a period of time exceeding 3\,ms. 
Initially, for the interval  $[1000 {\rm ms}, 1000.8 {\rm ms}]$, 
the backlog follows that of {\it PFC}, indicating that congestion control is not effective with reducing the backlog in this time interval. After $t=1000.8$\,ms, the backlog decreases at the link rate until it is completely cleared. 
The backlog curve for {\it DCQCN-noPFC} shows that, without the aid of PFC, the backlog increases for a duration of about 800\,$\mu$s and exceeds 50\,MB at its peak. 

To study the early stages of the experiment in more detail, 
Figure~\ref{fig:oneburst_31workers_delay1us_total400us} presents traffic rates and backlog for the initial 400\,$\mu$s. 
We first discuss {\it DCQCN-noPFC}. In Figure~\ref{fig:oneburst_31workers_delay1us_total400us_rate} we  observe that the aggregate traffic rate with {\it DCQCN-noPFC} decreases in a stepwise fashion. The rate reduction in each step is proportional to the rate, which reflects the multiplicative decrease in DCQCN. The first rate reduction is observed  within less than 10\,$\mu$s, at $t = 1000.01$\,ms. Subsequent rate reductions occur in intervals of 50\,$\mu$s, which is due to the 
time gap $T_{\rm gap}$ between two CNPs. 
At $t = 1000.4$\,ms, after 
eight rate reductions by the traffic senders, the aggregate arrivals 
still exceed the link capacity of the egress port to the destination. Therefore, the backlog continues to increase. 
The  traffic rates in Figure~\ref{fig:oneburst_31workers_delay1us_total400us_rate} 
for {\it DCQCN}, which includes PFC, are initially identical to {\it DCQCN-noPFC}. At about $t = 1000.13$\,ms, the PFC mechanisms is triggered, which drops the traffic rate to zero. 

\begin{figure}[!t]
\vspace{-0.2em}
\centering 
\subfloat[Total arrival rate.]{\includegraphics[width=0.48\textwidth]
{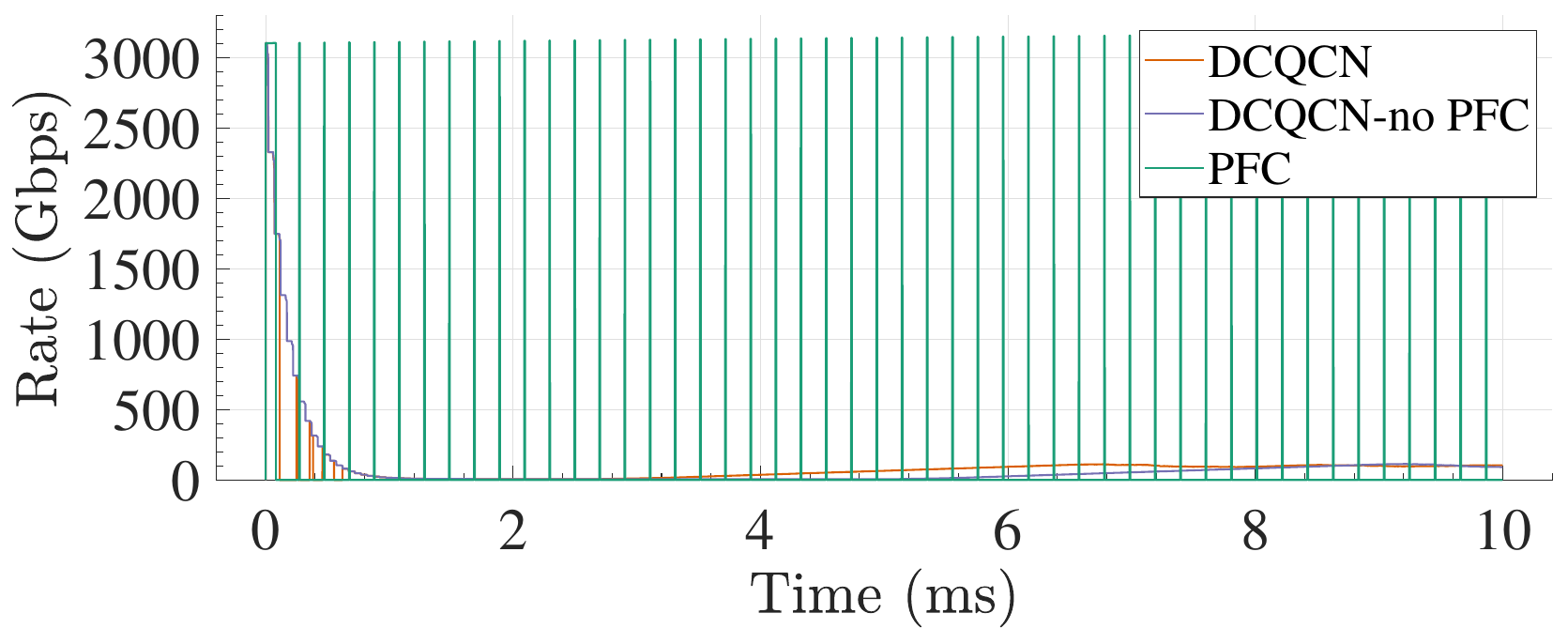}
\label{fig:AIMD_DCQCN_Rate_10ms}}
\subfloat[Backlog.]{\includegraphics[width=0.48\textwidth]{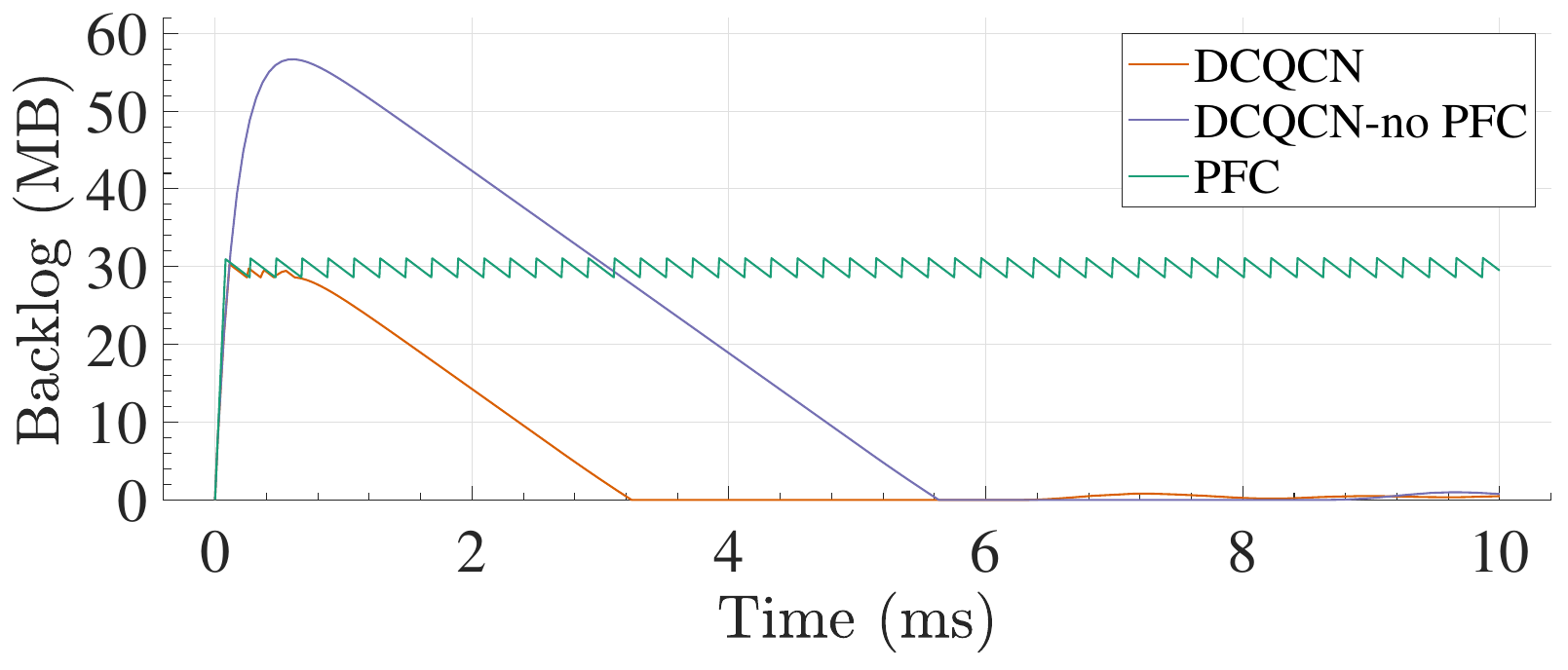}
\label{fig:AIMD_DCQCN_Backlog_10ms}}

\caption{Model-based characterization of burst arrivals (10~ms).}  
\label{fig:AIMD_DCQCN_10ms}
\centering 
\subfloat[Total arrival rate.]{\includegraphics[width=0.48\textwidth]
{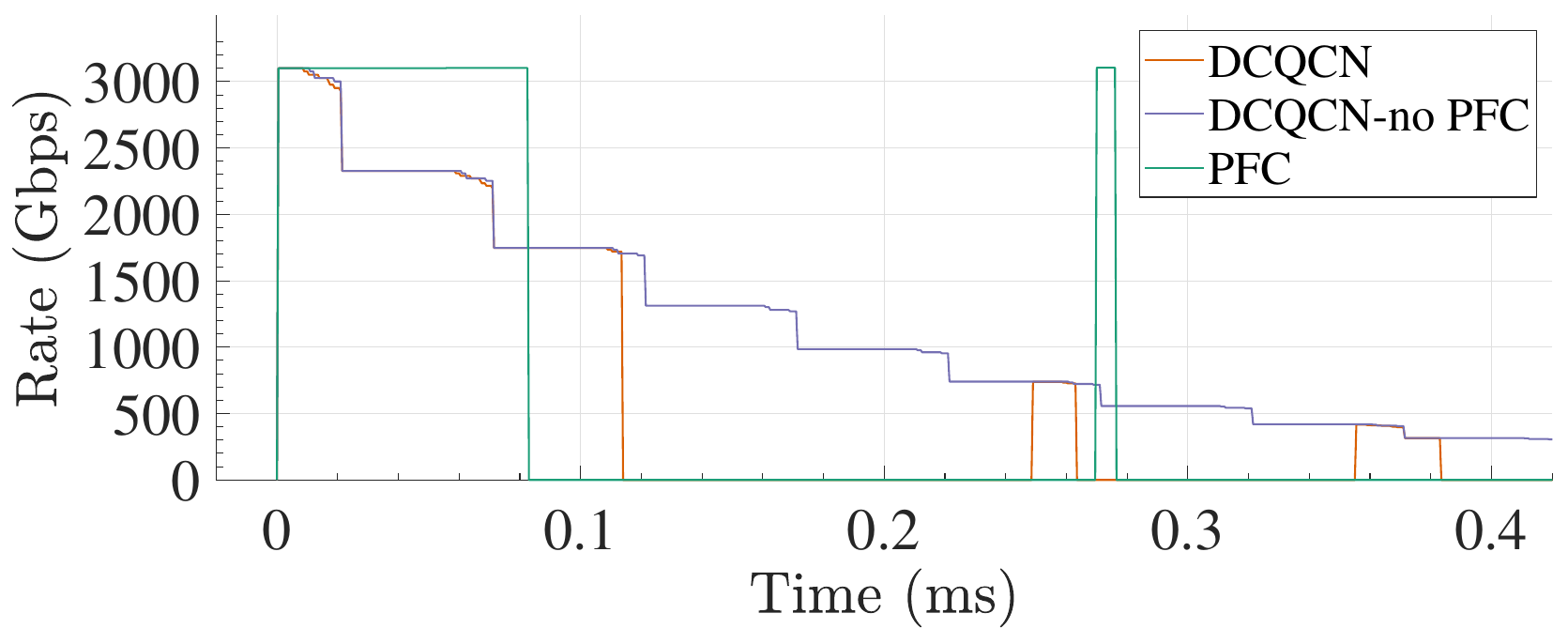}
\label{fig:AIMD_DCQCN_Rate_400us}}
\subfloat[Backlog.]{\includegraphics[width=0.48\textwidth]{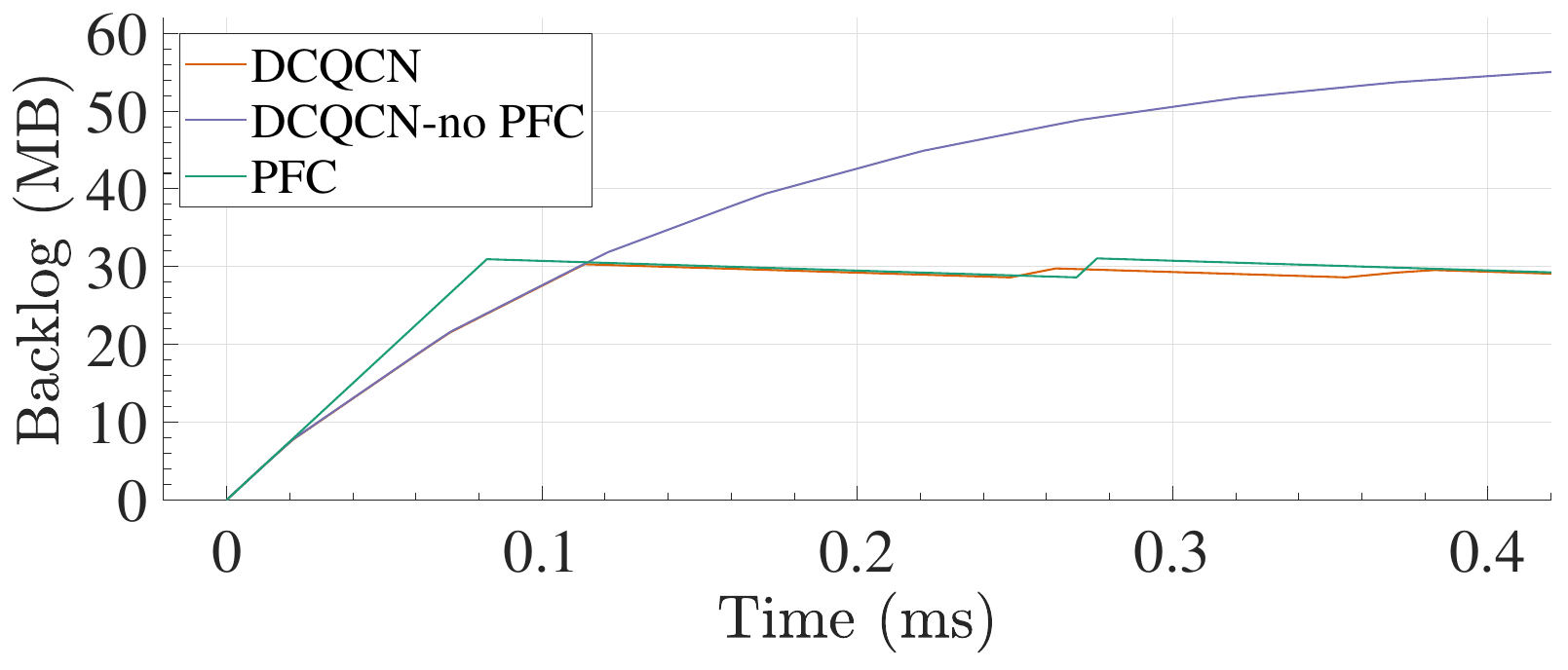}
\label{fig:AIMD_DCQCN_Backlog_400us}}

\caption{Model-based characterization of burst arrivals (400~$\mu$s).}   
\label{fig:AIMD_DCQCN_400us}
\end{figure} 

\subsection{Model-Based Analysis}
\label{subsec:dcqcn-model}

For the model-based analysis, we  consider a simplified version of DCQCN that focuses on the AIMD and ECN mechanisms. As a rate-based method, the model for DCQCN follows the analysis from~\S\ref{subsec:rate}. 
From Table~\ref{table:PFC-DCQCN-parameters} we obtain  $\alpha = 5$~Mbps  for the  additive increase, and $\Delta\tau_{inc}=55~\mu$s for the interval between increases. Our model does not 
account for hyper increase phase shown in Table~\ref{table:aimd}. For the multiplicative decrease, instead of accounting for a variable factor as used in DCQCN, we always apply a constant multiplicative factor $\beta = 0.75$, which is obtained empirically from 
simulation experiments. 
For ECN and PFC, our model follows the description in~\S\ref{subsec:events}. 
The ECN model accounts for  
RED with parameters given in Table~\ref{table:PFC-DCQCN-parameters}. The minimum distance between 
back-to-back congestion notifications is set to $\Delta\tau_{\rm ecn}= 50~\mu$s. 
The  PFC parameters are as  described above. 
Since PFC prevents packet losses, retransmission timeouts only play a role in the  {\it DCQCN-noPFC}  
scenario. We use a timeout value of $\tau_o=3$~ms in all scenarios. 

As arrival traffic, we consider simultaneous burst arrivals of 10~MB from 31 sources with  rate limits set to 100~Gbps at the start of the bursts. We set the service curve of the path server to $S(t) = \max\{ C \, t, 0\}$ where $C= 100$~Gbps. The feedback delay, $\Delta R$, is set to 4~$\mu$s, which corresponds to the total propagation round-trip delay.

Figures~\ref{fig:AIMD_DCQCN_10ms} and~\ref{fig:AIMD_DCQCN_400us} present the traffic rate and backlog obtained from our model, where Figures~\ref{fig:AIMD_DCQCN_400us} focuses on the 
first 400~$\mu$s. Traffic rates are computed by adding the admitted arrival functions $\hat{A}$ of all sources and then averaging over the same time intervals as in the simulations. 
A comparison with the simulation results shows that the network calculus model closely tracks the behavior of both DCQCN and PFC. The  -- sometimes subtle -- deviations are  due to  shortcuts in our DCQCN model. 
As one example, in Figure~\ref{fig:AIMD_DCQCN_Backlog_10ms} we see that the backlog for DCQCN  increases after an initial spike at around 7~ms. In the simulations~(Figure~\ref{fig:oneburst_31workers_delay1us_total10ms_backlog}) we observe the secondary spike much earlier at around 3~ms. Since we do not consider the hyper increase phase of DCQCN in our model, the traffic rates recover more slowly. This can be observed by comparing the transmission rates. In the simulations (Figure~\ref{fig:oneburst_31workers_delay1us_total10ms_rate}) we observe a non-linear jump to 300~Gbps at 3~ms,  whereas the model  (Figure~\ref{fig:AIMD_DCQCN_Rate_10ms}) shows  a linear increase until the 7~ms mark, when the maximum rate of 100~Gbps is reached.

\section{Related Work}
\label{sec:related}

The late 1990s and early 2000s were a golden era for analytical modeling of TCP congestion control, and many methods available today originate in that period. 
A seminal work by Kelly et al.~\cite{Kelly97}  presented  the first mathematical analysis of congestion control dynamics in a general network where rate allocations were formulated  in economic terms as a utility optimization. This and many works that followed~\cite{Low99,Low03,Low02,Srikant03,Book-srikant}  
derived  stability and convergence properties of  
various congestion control mechanisms and TCP variants

A different analysis approach from that era describes  
TCP dynamics by a coupled system of differential equations~\cite{Misra00} for tracking its 
transient behavior. With this, it is possible   to 
relate congestion control to classical control theoretic closed loop systems~\cite{Misra01,Misra02}. Until today, coupled differential equation systems are the prevalent method for  characterizing CAA dynamics~\cite{DCQCN-Sigcomm2015,DCTCPanalysis-11,fluid-BBR-22}. 
The above approaches  do not account for the  interactions between 
source traffic and congestion control, which is needed to study congestion control of intermittent but bursty traffic.  

Network calculus is an alternative approach for studying feedback systems. Agrawal et al.~\cite{Cruz99} derived a service curve for a window flow control system with a fixed window size. Baccelli and Hong~\cite{Baccelli00} expressed the feedback loop of TCP in a max-plus dioid algebra. Assuming fixed-sized packets, permanently backlogged sources, and excluding retransmissions they showed that  TCP satisfies max-plus linearity.\footnote{Max-plus and min-plus linear systems are dual to each other~\cite{Liebeherr17}.} Kim and Hou~\cite{Hou04} showed that the rate limits imposed by AIMD can be viewed as a piece-wise linear shaper. 
A recent study~\cite{Arun_2021} applied  
techniques inspired by network calculus to study scenarios where  CCAs fail. Specific loss scenarios are expressed  in terms of constraints on  arrivals and departures to create a satisfiability problem. In~\cite{Arun_2021}, the network is represented as a generalized token bucket, referred to as path server, which  determines the available bandwidth to a traffic flow. The network model of our paper is derived from the path server model and shares many of its assumptions. 

\section{Conclusions}

We have developed a modular model-based approach to the analysis of CCAs that can account for bursty traffic sources. Using a path server network model, we characterized rate-based and window-based congestion control algorithms within the framework of the network calculus. 
By performing an analysis within time intervals where the network satisfies  linearity conditions and appropriately shifting the reference coordinate system, we were able to conduct a linear analysis of an overall non-linear network. 
We showed that our model-based characterization can capture convergence toward fairness of a CCA. The network calculus characterization of 
congestion control provides an alternative tool for studying the dynamics of CCA, which can achieve a high degree of accuracy. This was demonstrated in a  case study of burst transmissions in 
a data center network which deploy PFC and DCQCN. 

\newpage 


\clearpage

\begin{appendices}
\section{Algorithm for rate-based congestion control}
\label{sec:rate-algorithm}
Algorithm~\ref{alg:AIMD-departures} summarizes the computations for a rate-based CCA with an AIMD scheme.  The algorithm takes as input an arrival function in the range $[0,t_{\rm end}]$. 
The service curve of the path server is a rate function with initial rate $r_o$.
We can distinguish five phases: (1) initial calculation of $\hat{A}$ and $D$ with the input parameters (steps 1--4); (2)  determination of the next congestion event (steps~6--8), (3) adjustment of the  rate limit based on the type of congestion event (steps 10--12 and 17--18),   (4) updating the arrival and admitted arrival functions in a shifted  coordinate system (steps 12--15 and 19--22), and finally (5) updating the departure function (steps 24--25). Steps (2)--(5) are repeated until no more congestion events are found.


\begin{algorithm}[!h]
\SetAlgoLined
\DontPrintSemicolon
\BlankLine
\KwIn{~~~~~$A, S,\alpha,\beta,\tau_o,r_o,\Delta R, \Delta \tau_{\rm ai},t_{\rm end} $}
\KwOut{~~$D, \hat{A}$}
\BlankLine
$r\gets r_o$ \, , $t^{\rm prev}_{\rm TO} \gets 0$ \, , $t_{\rm AI}  \gets 0$\;
$S_{r}(t) \gets \max\{rt,0\}$, for $t \in [0,t_{\rm end}]$\;
$\hat{A} (t) \gets  A \otimes S_{r} (t)$ , for $t \in [0,t_{\rm end}]$ \;
$D (t) \gets  \hat{A} \otimes S (t)$ , for $t \in [0,t_{\rm end}]$ \;

  \Repeat{$t_{\rm event} \geq t_{\rm end}$}{

	$t_{\rm AI} \gets \min \left\{  t_{\rm AI} + \Delta \tau_{\rm ai}, t_{\rm end} \right\}$\;
	
	$t^{\rm next}_{\rm TO} \gets \min \left\{ t>t^{\rm next}_{\rm TO} \;|\; D(t-\Delta R)<\hat{A}(t - \tau_o)  \right\}$\;
 
	$t_{\rm event} \gets \min \left\{ t^{\rm next}_{\rm TO}, t_{\rm AI}\right\}$\;
    \BlankLine
	\eIf{ $t_{\rm event} = t^{\rm next}_{\rm TO}$ }
	{
        \BlankLine
		$t^{\rm prev}_{\rm TO} \gets t^{\rm next}_{\rm TO}$\; 
		$r \gets \beta r$\;
        $S_{r}(t) \gets\max\{rt,0\}$, for $t \in [0,t_{\rm end}-t_{\rm event}]$\;
		$A^{\rm rem}(t) \gets A(t+t_{\rm event})-D(t_{\rm event}-\Delta R)$, for $t \in [0,t_{\rm end}-t_{\rm event}]$ \;
  
		$\hat{A}^{\rm rem}(t)\gets A^{\rm rem}\otimes S_{r} (t)$, for $t \in [0,t_{\rm end}-t_{\rm event}]$ \;
  
		$\hat{A}(t)\gets \hat{A}^{\rm rem}(t-t_{\rm event})+D(t_{\rm event}-\Delta R)$,  for $t \in [t_{\rm event},t_{\rm end}]$ \;
		
	}
	{
        
		$r\gets r + \alpha$\;
        $S_{r}(t) \gets \max\{rt,0\}$, for $t \in [0,t_{\rm end}-t_{\rm event}]$\;
		$A^{\rm rem}(t) \gets A(t+t_{\rm event})-A(t_{\rm event})$, \,for $t \in [0,t_{\rm end}-t_{\rm event}]$ \;
  
		$\hat{A}^{\rm rem}(t)\gets A^{\rm rem}\otimes S_{r} (t)$, \,for $t \in [0,t_{\rm end}-t_{\rm event}]$ \;
  
		$\hat{A}(t) \gets \hat{A}^{\rm rem}(t-t_{\rm event})+A(t_{\rm event})$, \,for $t \in [t_{\rm event},t_{\rm end}]$ \;
  
		$\hat{A}^{\rm rem}(t)\gets \hat{A}(t+t^{\rm prev}_{\rm TO})-D(t^{\rm prev}_{\rm TO})$, \,for $t \in [0,t_{\rm end}-t^{\rm prev}_{\rm TO}]$ \;
	}	
	
	$D^{\rm rem}(t) \gets \hat{A}^{\rm rem} \otimes S (t)$, \,for $t \in [0,t_{\rm end}-t^{\rm prev}_{\rm TO}]$ \;
	$D(t) \gets D^{\rm rem}(t-t^{\rm prev}_{\rm TO}) +D(t^{\rm prev}_{\rm TO}-\Delta R)$, \,for $t \in [t^{\rm prev}_{\rm TO},t_{\rm end}]$ 

  }
\BlankLine
\caption{Network calculus computation of rate-based AIMD congestion control.}
\label{alg:AIMD-departures}
\end{algorithm}

\section{Algorithm for window-based congestion control}
\label{sec:window-algorithm}

Algorithm \ref{alg:TCP-departures} presents the computations for a 
window-based flow control, which is similar to TCP Tahoe, for an 
arrival function $A(t)$ for $t \le t_{\rm end}$. 
 Like Algorithm \ref{alg:AIMD-departures}, the algorithm updates $\hat{A}$ and $D$ in iterations depending on encountered congestion events. Steps 11--14 and steps 18--20 show the updates of $\hat{A}$ after an update of the window size.

\begin{algorithm}[h]
\SetAlgoLined
\DontPrintSemicolon
\BlankLine
\KwIn{~~~~~$A, S, \alpha,\beta, \tau_o,\Delta R, W_o , {w_{th}}_o \text{(initial slow start threshold)},t_{\rm end} $}
\KwOut{~~$D, \hat{A}$}
\BlankLine
$W\gets W_o$\, , $t^{\rm prev}_{\rm TO} \gets 0$ \;
$w_{th}\gets {w_{th}}_o$\;
$S_{W}(t) \gets W$, for $t \in [0,t_{\rm end}]$\;
$\hat{A}(t) \gets  \hat{A} \otimes S_W(t) $, for $t \in [0,t_{\rm end}]$ \;
$D(t) \gets  \hat{A} \otimes S(t)$, for $t \in [0,t_{\rm end}]$ \;
  \Repeat{$t_{\rm event} \geq t_{\rm end}$}{

	$t_{W} \gets \min \left\{ t \;|\; D(t-\Delta R)>W  \right\}$\;
	
	$t^{\rm next}_{\rm TO} \gets \min \left\{ t \;|\; D(t-\Delta R)<\hat{A}(t - \tau_o)  \right\}$\;

   $t_{\rm event} \gets \min \left\{ t^{\rm next}_{\rm TO}, t_{W}\right\}$\;
	
	\BlankLine
	\eIf{ $t_{\rm event}=t^{\rm next}_{\rm TO}$ }
	{
        \BlankLine
		$t^{\rm prev}_{\rm TO} \gets t^{\rm next}_{\rm TO}$ \; 
		$w_{th}\gets \beta W$\;
		$W \gets W_o$\;
		$A^{\rm rem}(t) \gets A(t+t_{\rm event})-D(t_{\rm event}-\Delta R)$, for $t \in [0,t_{\rm end}-t_{\rm event}]$ \;
        $S_{W}(t) \gets W$, for $t \in [0,t_{\rm end}-t_{\rm event}]$\;
		$\hat{A}^{\rm rem}(t)\gets A^{\rm rem} \otimes S_W (t)$, for $t \in [0,t_{\rm end}-t_{\rm event}]$ \;
		
		$\hat{A}(t)\gets \hat{A}^{\rm rem}(t-t_{\rm event})+D(t_{\rm event}-\Delta R)$, \,for $t \in [t_{\rm event},t_{\rm end}]$ \;
		
	}
	{
		Update $W$ according to Eq.~\eqref{windowFunction}\;
		$A^{\rm rem}(t) \gets A(t+t_{\rm event})-A(t_{\rm event})$, for $t \in [0,t_{\rm end}-t_{\rm event}]$ \;
        $S_{W}(t) \gets W$, for $t \in [0,t_{\rm end}-t_{\rm event}]$\;
		$\hat{A}^{\rm rem}(t)\gets A^{\rm rem} \otimes S_W (t)$, for $t \in [0,t_{\rm end}-t_{\rm event}]$ \;
  
		$\hat{A}(t) \gets \hat{A}^{\rm rem}(t-t_{\rm event})+A(t_{\rm event})$, \,for $t \in [t_{\rm event},t_{\rm end}]$ \;
  
		$\hat{A}^{\rm rem}(t)\gets \hat{A}(t+t^{\rm prev}_{\rm TO})-D(t^{\rm prev}_{\rm TO})$, \,for $t \in [0,t_{\rm end}-t^{\rm prev}_{\rm TO}]$ \;
		
	}
	
	$D^{\rm rem} \gets \hat{A}^{\rm rem} \otimes S (t)$, \,for $t \in [0,t_{\rm end}-t^{\rm prev}_{\rm TO}]$ \;
	$D(t) \gets D^{\rm rem}(t-t^{\rm prev}_{\rm TO}) +D(t^{\rm prev}_{\rm TO}-\Delta R)$, \,for $t \in [t^{\rm prev}_{\rm TO},t_{\rm end}]$\;

  }
\BlankLine
\caption{Network calculus computation of window-based AIMD congestion control.}
\label{alg:TCP-departures}
\end{algorithm}

\newpage 
\section{Algorithm for TCP Vegas}
\label{sec:vegas-algorithm}
Algorithm \ref{alg:TCP-Vegas} summarizes the computations  for  TCP Vegas. 
Whenever the current 
RTT estimate $\textit{minRTT}$ is updated (lines 6-7), we update $W$ (lines 8-25), along with $\hat{A}$~and~$D$ (lines 26-30). The algorithm does not account for timeouts and retransmissions. We indicate slow start using $\text{Flag}_{\rm slow start}$ and $\text{Flag}_{\rm skip}$ ensures updates only occur every other RTT in this phase. We use $w_{\rm low}$ and $w_{\rm high}$ to denote the $\alpha$ and $\beta$ thresholds in \cite{Vegas,vegasVernon},  respectively. The definitions for $\gamma, \textit{baseRTT}$ and $ \textit{diff}$ are identical to \cite{Vegas,vegasVernon}. 
\begin{algorithm}[h]
\SetAlgoLined
\DontPrintSemicolon
\BlankLine
\KwIn{~~~~~$A, S, \Delta R, W_o ,t_{\rm end}, I_{\max}, w_{\rm low},w_{\rm high}, \gamma $}
\KwOut{~~$D, \hat{A}$}
\BlankLine
$W\gets W_o$\, , $t_{\rm event} \gets 0$, $\text{Flag}_{\rm slow start} \gets 1$ ,  $\text{Flag}_{\rm skip} \gets 0$ \;
$\textit{baseRTT} \gets \Delta R$, $\textit{minRTT} \gets \Delta R$\;
$\hat{A}(t) \gets  \hat{A} \otimes S_W(t) $, for $t \in [0,t_{\rm end}]$ \;
$D(t) \gets  \hat{A} \otimes S(t)$, for $t \in [0,t_{\rm end}]$ \;
  \Repeat{$t_{\rm event} \geq t_{\rm end}$}{

   $t_{\rm event} \gets t_{\rm event} +\textit{minRTT}$\;

   $\textit{minRTT} \gets RTT(t_{\rm event})$ according to Eq.~\eqref{eq:RTT}\;
    
     $\textit{diff}\gets \textit{baseRTT}\cdot(\frac{W}{\textit{baseRTT}}-\frac{W}{\textit{minRTT}})$\;
	\eIf{$\text{Flag}_{\rm slow start} = 1$}
    {
    \BlankLine
    \eIf{$\textit{diff}>\gamma$}
        {
            $W \gets  W\cdot \frac{\textit{baseRTT}}{\textit{minRTT}}$\;
            $\text{Flag}_{\rm slow start} \gets 0$\;
        }
        {
            \If{$\text{Flag}_{\rm skip} =0$} 
            {
                $W \gets 2 \cdot W$\;
            }
            $\text{Flag}_{\rm skip} \gets (1+\text{Flag}_{\rm skip})\mod 2$\;
        }
    }
    {
        \uIf{$\textit{diff}<w_{\rm low}$}
        {
             $W \gets  W+I_{\max}$\;
        }
        \ElseIf{$\textit{diff}>w_{\rm high}$}
        {
             $W \gets W-I_{\max}$\;
        }
    }
	$A^{\rm rem}(t) \gets A(t+t_{\rm event})-A(t_{\rm event})$, for $t \in [0,t_{\rm end}-t_{\rm event}]$ \;
    $\hat{W} (t) \gets \max \left\{0,W-(\hat{A}(t_{\rm event}) -D(t+t_{\rm event} -\Delta R)) \right\} $, for $t \in [0,t_{\rm end}-t_{\rm event}]$ \;
	$\hat{A}^{\rm rem}(t)\gets \min \left\{A^{\rm rem}(t), \hat{W}(t)\right\}$, for $t \in [0,t_{\rm end}-t_{\rm event}]$ \;
  
	$\hat{A}(t) \gets \hat{A}^{\rm rem}(t-t_{\rm event})+A(t_{\rm event})$, \,for $t \in [t_{\rm event},t_{\rm end}]$ \;
  
	$D(t) \gets  \hat{A} \otimes S(t)$, for $t \in [0,t_{\rm end}]$ \;
  }
\BlankLine
\caption{Network calculus computation of TCP Vegas without timeouts.}
\label{alg:TCP-Vegas}
\end{algorithm}

\clearpage
\section{Proofs}
\label{sec:proofs}

\subsection{Proof of Lemma~\ref{lemma:aggregate_fifo_property}}
	To show Eq.~\eqref{eq:aggregate_fifo_property:part_imply_sum}, select a flow $i \in {\mathcal N}$ and an interval $[t', t]$ with~$\hat{A}_i(t')< D_i(t)$.
	From the FIFO property, we can conclude that every flow~$j\in X$ satisfies $\hat{A}_j(t') \le D_j(t)$. 
	Summing up over all flows in $X$ yields
	\begin{equation*}
        \sum_{j\in X} \hat{A}_j(t') \le \sum_{j\in X} D_j(t)\,.
	\end{equation*}
    For Eq.~\eqref{eq:aggregate_fifo_property:sum_imply_part}, we prove its converse by selecting a flow~$i\in{\mathcal N}$ and an interval $[t', t]$ with~$\hat{A}_i(t') > D_i(t)$.
    Then, due to the FIFO property, every flow~$j\in X$ satisfies $\hat{A}_j(t')\ge D_j(t)$.
	In this case, summing up over all flows in~$X$ yields
	\begin{equation*}
        \sum_{j\in X}\hat{A}_j(t') \ge \sum_{j\in X}D_j(t)\,.
	\end{equation*}
\hfill \qedsymbol{}

\subsection{Proof of Lemma~\ref{lemma:strict_aggregate_fifo_property}}
 Consider a set of flows that satisfies Eq.~\eqref{eq:strict_aggregate_fifo_property:antecedent}.
    Suppose that there is some flow~$i\in X$ where~$\hat{A_i}(t') < D_i(t)$.
    Due to the FIFO property, it holds that every flow~$j\in X$ satisfies~$\hat{A}_j(t')\le D_j(t)$.
    Summing up over all flows in~$X$ yields
    \begin{equation*}
        \sum_{j\in X} \hat{A}_j(t') < \sum_{j\in X} D_j(t),
    \end{equation*}
    which contradicts Eq.~\eqref{eq:strict_aggregate_fifo_property:antecedent}.
    So every flow~$i\in X$ satisfies~$\hat{A_i}(t') \ge D_i(t)$.

    Similarly, suppose that some flow~$i\in X$ satisfies~$\hat{A_i}(t') > D_i(t)$.
    Due to the FIFO property, every other flow~$j\in X$ satisfies~$\hat{A}_j(t')\ge D_j(t)$.
    Summing up over all flows in~$X$ yields
    \begin{equation*}
        \sum_{j\in X} \hat{A}_j(t') < \sum_{j\in X} D_j(t),
    \end{equation*}
    which contradicts Eq.~\eqref{eq:strict_aggregate_fifo_property:antecedent}, and so~$\hat{A_i}(t') = D_i(t)$ for all~$i\in X$. \hfill \qedsymbol{}

\subsection{Proof of Theorem~\ref{thm:split_fifo}}

The proof uses a property of the upper pseudo-inverse~\cite[p.~66, (P4)]{Liebeherr17}, which states that every left-continuous function $F$ satisfies for every~$y\in\mathbb R$,
	\begin{equation*}
		F(F^{\uparrow}(y)) \le y < F(F^\uparrow(y) + \varepsilon)\,.
	\end{equation*}
	Setting $F = \hat{A}$ and $y = D(t)$, we obtain
	\begin{equation*}
		\hat{A}(\hat{A}^{\uparrow}(D(t))) \le D(t) < \hat{A}(\hat{A}^{\uparrow}(D(t)) + \varepsilon)\,.
	\end{equation*}
	We then apply Lemma~\ref{lemma:aggregate_fifo_property} to both inequalities in the above equation.
    For the right inequality, we set~$X = {\mathcal N}$ and~$t' = \hat{A}^{\uparrow}(D(t)) + \varepsilon$ to the converse of Eq.~\eqref{eq:aggregate_fifo_property:part_imply_sum} to obtain 
	\begin{equation*}
		D_i(t) < \hat{A}_i(\hat{A}^{\uparrow}(D(t)) + \varepsilon)\,.
	\end{equation*}
    For the left inequality, we split it into two cases.
    If~$\hat{A}(\hat{A}^{\uparrow}(D(t))) < D(t)\,$, we apply Eq.~\eqref{eq:aggregate_fifo_property:sum_imply_part} with~$X = {\mathcal N}$ and~$t' = \hat{A}^{\uparrow}(D(t))$ to arrive at
	\begin{equation}
        \label{eq:split_fifo:left_ineq}
		\hat{A}_i(\hat{A}^{\uparrow}(D(t))) \le D_i(t)\,. 
	\end{equation}
    Otherwise,~$\hat{A}(\hat{A}^{\uparrow}(D(t))) = D(t)\,$, and we can apply Lemma~\ref{lemma:strict_aggregate_fifo_property} to arrive at Eq.~\eqref{eq:split_fifo:left_ineq}.\hfill \qedsymbol{}


\end{appendices}

\end{document}